\documentclass[showpacs,preprintnumbers,amsmath,amssymb]{revtex4}

 \usepackage{graphics,epsfig,subfigure}

\usepackage{amssymb,amsmath,bm}
\usepackage{cases}

\usepackage{verbatim}
\usepackage{amstext}
\usepackage{color,changes,ulem}
\usepackage{array}
\usepackage{multirow}
\newcommand\beq{\begin{equation}}
\newcommand\eeq{\end{equation}}
\newcommand\beqn{\begin{eqnarray}}
\newcommand\eeqn{\end{eqnarray}}

\begin{document}
\renewcommand{\baselinestretch}{1.3}

\title{Phase transition and microstructures of five-dimensional charged Gauss-Bonnet-AdS black holes in the grand canonical ensemble}

\author{Run Zhou,
    Yu-Xiao Liu,
	Shao-Wen Wei \footnote{weishw@lzu.edu.cn}}

\affiliation{Institute of Theoretical Physics $\&$ Research Center of Gravitation, Lanzhou University, Lanzhou 730000, People's Republic of China,\\
Joint Research Center for Physics, Lanzhou University and Qinghai Normal University, Lanzhou 730000 and Xining 810000, China}

\begin{abstract}
In this paper, we study the small-large black hole phase transition and construct the Ruppeiner geometry for the five-dimensional charged Gauss-Bonnet-AdS black hole in the grand canonical ensemble. By making use of the equal area law, we obtain the analytical coexistence curve of the small and large black holes. Then the phase diagrams are examined. We also calculate the change of the thermodynamic volume during the small-large phase transition, which indicates that there exists a sudden change among the black hole microstructures. The corresponding normalized scalar curvature of the Ruppeiner geometry is also calculated. Combing with the empirical observation of scalar curvature, we find that for low electric potential, the attractive interaction dominates among the microstructures, while a high electric potential produces repulsive interactions. In the reduced parameter space, we observe that only attractive interaction is allowed when the coexistence region is excluded. The normalized scalar curvature also admits a critical exponent 2 and a universal constant $-\frac{1}{8}$. In particular, the value of the normalized scalar curvature keeps the same along the coexistence small and large black hole curves. So in the grand canonical ensemble, the interaction can keep constant at the phase transition where the black hole microstructures change. These results disclose the intriguing microstructures for the charged AdS black hole in the Gauss-Bonnet gravity.
\end{abstract}



\maketitle

\section{Introduction}

Since the establishment of the four thermodynamic laws of black holes, thermodynamics and phase transition continue to be one of the increasingly active areas in black hole physics. Hawking and Page observed a phase transition between the pure thermal radiation and stable large black hole \cite{Hawking1983b}. Adopting the AdS/CFT correspondence \cite{Maldacena,Gubser,Witten}, such phase transition was interpreted as the  confinement/deconfinement phase transition of a gauge field \cite{Witten2}. Therefore black hole phase transition attracted much attention. Interestingly, the van der Waals (VdW) like phase transition was observed in charged and rotating black holes in AdS space \cite{Chamblin,Chamblin2,Caldarelli}.

Recent study of black hole thermodynamics and phase transitions was accompanied by the understanding of the cosmological constant. As early as 1984, Brown and Teitelboim first proposed that the cosmological constant can be treated as a dynamic variable \cite{Henneaux1984,Teitelboim1985}. This idea was put forward in Refs. \cite{Creighton1995,Padmanabhan2002}. In 2009, Kastor, Ray and Traschen made a significant progress that the cosmological constant was interpreted as the pressure of the black hole system \cite{Kastor2009d}. Then it was found that the first law of thermodynamics is consistent with the Smarr relation. Meanwhile, the black hole mass will be regarded as the enthalpy rather the internal energy of the black hole system, and the thermodynamic volume will be found by using the first law \cite{Dolan00,Cvetic}. The precise analogy between the small-large black hole phase transition and liquid-gas phase transition was completed later by Kubiznak and Mann \cite{Kubiznak2012b}. Subsequently, more new black hole phase transitions and phase structures were observed, such as the reentrant phase transitions, isolated critical points, triple points, and superfluid black hole phases \cite{Gunasekaran,Altamirano,Mann,Frassino,Wei0,Kostouki,Wei1,Hennigar,ZouYue,
Hendi3,Hendi4,Momeni,Chakraborty,Weisw} (for a recent review see \cite{Teo} and references therein).

As we know, macroscopic thermodynamics of a system originates from its microstructures. Similarly, we believe that this also holds for the black hole systems. However, it is well known that the entropy of a black hole is proportional to the area of the event horizon rather than its volume. The study of this property will help us to deeply understand the underlying black hole microstructures. There are several different approaches to derive the Beckenstein-Hawking entropy area formula. String theory provides a preliminary calculation by counting the number of states of a weakly coupled D-brane system \cite{Vafa}. Fuzzball theory also shows an understanding of the black hole microstructures \cite{Lunin,Mathur}. Based on the Cardy formula \cite{Cardy1986}, the entropy area formula can also be obtained \cite{Banados1992}.

Among the study of the black hole microstructures, the Ruppeiner geometry \cite{Ruppeiner1979,Ruppeiner1995} provides a powerful tool. By using the empirical observations of the Ruppeiner geometry, the interactions dominated among the microstructures can be indicated from the scalar curvature. The positive or negative scalar curvature corresponds to repulsive or attractive interaction. Combining with the small-large black hole phase transition, we constructed the Ruppeiner geometry for the charged AdS black holes by assuming that black hole is constituted by some unknown molecules. Then the properties of the black hole microstructures were uncovered \cite{Weiw,Wei:2020cqn}. This approach has also been generalized to different black hole backgrounds \cite{Dehyadegari,Zangeneh:2016fhy,Moumni,Deng,Sheykhi,Miao,Miao2018,Miao2,Miao3,
Li,Yang:2018ixs,Chen,Guo,Du,Sheykhi:2019vzb,Xuz,GhoshBhamidipati}.

Since the critical phenomenon was not observed in our previous paper \cite{Weiw}, we reconsidered the Ruppeiner geometry for the charged AdS black holes. After constructing the new geometry, we found that besides the attractive interaction, the repulsive interaction can also dominate between the black hole molecules. The critical exponents and universal constant were also studied in detail  \cite{Wei2019}. Subsequently, this novel approach was applied to other black holes in AdS space \cite{Wei2019d,Kumara2020,Bairagya:2020dtl,Yerra:2020oph,Wu:2020fij,Dehyadegari:2020ebz,Wei2020d,Vaid,Rizwan,Mansoori,Kumara,Mannw,Dehyadegariw}.

In particular, this approach was also applied to the five-dimensional neutral Gauss-Bonnet (GB) AdS black hole \cite{Wei2020a}. Employing the analytical coexistence curve, we observed that only the attractive interaction dominates among the black hole molecules. Another intriguing result shows that among the small-large black hole phase transition, the microstructures gain a huge change, while the interactions keep unchanged. This gives a first example that the change of the microstructures has no influence on the microscopic interactions. It also uncovers the characteristic properties of the GB gravity.

Since the charge is absent in the study of \cite{Wei2020a}, we here would like to consider the black hole microstructures when the charge is included, and to study whether the result of \cite{Wei2020a} holds. We deal with the five-dimensional charged GB-AdS black holes. After constructing the equal area law, we find that there exists an analytical coexistence curve in the grand ensemble. It is very helpful for exactly understanding the characteristic properties of the GB gravity.

The structure of this paper is as follows. In Sec. \ref{null}, we briefly review the thermodynamic properties of the five-dimensional charged GB-AdS black hole. In Sec. \ref{phase}, we construct the equal area law in the $P$-$V$ plane, and obtain the analytical coexistence curve of the small-large black hole phase transition. The Critical exponents are also calculated. The Ruppeiner geometry is constructed in Sec. \ref{Ruup}. By calculating the corresponding scalar curvature, the properties of the black hole microstructures are investigated, and the critical phenomena of the normalized scalar curvature are analyzed. Finally, the conclusions and discussions are given in Sec. \ref{Conclusion}. Throughout this paper, we adopted the units $\hbar$=$c$=$k_B$=$G_5$=1.

\section{Thermodynamics of five-dimensional charged Gauss-Bonnet-AdS black holes}
\label{null}

In this section, we present a brief review of the thermodynamics of the five-dimensional charged GB-AdS black hole \cite{Cai2013,Zou2014c,Belhaj2015}. This black hole solution is described by the following action
\begin{eqnarray}
 &S=\int d^{5}x\sqrt{-g}
 \left(\frac{1}{16\pi G_{5}}(\mathcal{R}
    -2\Lambda
    +\alpha_{\text{GB}}\mathcal{L}_{\text{GB}})
  -\mathcal{L}_{\text{matter}}\right), \label{action}
\end{eqnarray}
where
\begin{eqnarray}
   \mathcal{L}_{\text{GB}}~~&=&\mathcal{R}_{\mu\nu\gamma\delta}
   \mathcal{R}^{\mu\nu\gamma\delta}
                    -4\mathcal{R}_{\mu\nu}\mathcal{R}^{\mu\nu}+\mathcal{R}^{2},\\
 \mathcal{L}_{\text{matter}}&=&4\pi \mathcal{F}_{\mu\nu}\mathcal{F}^{\mu\nu}.
\end{eqnarray}
The Maxwell field strength is defined as
$\mathcal{F}_{\mu\nu}=\partial_{\mu}\mathcal{A}_{\nu}-\partial_{\nu}\mathcal{A}_{\mu}$
with $\mathcal{A}_{\mu}$ the vector potential. The line element is given by
\begin{eqnarray}
 d s^{2}=-f(r) d t^{2}+f^{-1}(r) d r^{2}+r^{2} \left(d \theta_{1}^{2}+\sin ^{2} \theta_{1}\left(d \theta_{2}^{2}+\sin ^{2} \theta_{2} d \theta_{3}^{2}\right)\right),
\end{eqnarray}
with
\begin{eqnarray}
f(r)=1+\frac{r^{2}}{2 \alpha}\left(1-\sqrt{1+\frac{32  \alpha M}{ 3\pi r^{4}}-\frac{\alpha Q^{2}}{3 r^{6}}-\frac{16 \pi \alpha P}{3}}\right),\label{fr}
\end{eqnarray}
where the parameters $M$ and $Q$ are the black hole mass and charge, respectively. Pressure $P$ is related to the cosmological constant as $P=-\frac{\Lambda}{8 \pi}$ and $\alpha = 2 \alpha_{GB}$ is the GB coupling. Here we only consider the case of positive $\alpha_{GB}$.

The radius $r_\text{h}$ of the black hole event horizon is the largest root of $f(r_\text{h})=0$. In terms of $r_\text{h}$, the black hole mass reads
\begin{eqnarray}
M= \frac{  4 \pi \left(3 \alpha r_\text{h}^2+4 \pi  Q r_\text{h}^6+3 r_\text{h}^4\right)+\pi Q^2}{32 r_\text{h}^2}.\label{ent}
\end{eqnarray}
The corresponding Hawking temperature is
\begin{eqnarray}
T=\frac{1}{4 \pi} f^{\prime}\left(r_\text{h}\right)=-\frac{-32 \pi  P r_\text{h}^6+Q^2-12 r_\text{h}^4}{48 \pi  \alpha r_\text{h}^3+24 \pi  r_\text{h}^5}.\label{tem}
\end{eqnarray}
In the extended phase space, the black hole mass acts as the enthalpy $H=M$ of the system. The entropy $S$, thermodynamic volume $V$ and electric potential $\Phi$ can be calculated as
\begin{eqnarray}
 &&S=\int T^{-1}\left(\frac{\partial H}{\partial r}\right)_{Q, P} d r=3 \alpha \pi^{2} r_\text{h}+\frac{1}{2} \pi^{2} r_\text{h}^{3}\label{entropy},\\
 &&V=\left(\frac{\partial H}{\partial P}\right)_{S, Q}=\frac{\pi^{2} r_\text{h}^{4}}{2}\label{volume},\\
 &&\Phi=\left(\frac{\partial H}{\partial Q}\right)_{S, P}=\frac{\pi Q}{16 r_\text{h}^{2}}.\label{ephi}
\end{eqnarray}
In terms of $\Phi$, the enthalpy $H$ and temperature $T$ can be rewritten as
\begin{eqnarray}
H&=&\frac{1}{8} \pi\left(3 \alpha+3 r_\text{h}^{2}+4 P \pi r_\text{h}^{4}\right)+\frac{8 r_\text{h}^{2} \Phi^{2}}{\pi},\\
T&=&\frac{r_\text{h}\left(\pi^{2}\left(3+8 P \pi r_\text{h}^{2}\right)-64 \Phi^{2}\right)}{6 \pi^{3}\left(2 \alpha+r_\text{h}^{2}\right)}.\label{temperature}
\end{eqnarray}
It is easy to check that those thermodynamic quantities satisfy the following first law and Smarr relation
\begin{eqnarray}
 dH&=&TdS+VdP+\mathcal{A}d\alpha,\\
 2H&=&3TS-2PV+2\mathcal{A}\alpha,
\end{eqnarray}
where $\mathcal{A}$ is the conjugate quantity to the GB coupling $\alpha$, which is given by
\begin{eqnarray}
 \mathcal{A}=\frac{\pi  \left(-32 \pi  P r_h^6+6 \alpha
   r_h^2-9 r_h^4+Q^2\right)}{8 r_h^2 \left(2
   \alpha +r_h^2\right)}.
\end{eqnarray}
From Eq. (\ref{temperature}), we can express $P$ as a function of $T$, $V$, $\alpha$, and $\Phi$
\begin{eqnarray}
P=\frac{\left(64 \Phi^{2}-3 \pi^{2}\right) \sqrt{2}}{16 \pi^{2}V^{\frac{1}{2}}} +\frac{3 \times 2^{\frac{3}{4}} \pi^{\frac{1}{2}} T}{8V^{\frac{1}{4}}} +\frac{3 \times 2^{\frac{1}{4}} \pi^{\frac{3}{2}} T \alpha}{4V^{\frac{3}{4}}}. \label{statep}
\end{eqnarray}
The critical point is determined by the following condition
\begin{eqnarray}
(\partial_{V} P)_{T, \alpha, \Phi}=(\partial_{V,V} P)_{T, \alpha, \Phi}=0,
\end{eqnarray}
which gives
\begin{eqnarray}
T_\text{c}=\frac{3 \pi^{2}-64 \Phi^{2}}{6 \sqrt{6} \sqrt{\alpha} \pi^{3}},\quad P_\text{c}=\frac{3 \pi^{2}-64 \Phi^{2}}{144 \alpha \pi^{3}},\quad V_\text{c}=18 \alpha^{2} \pi^{2}.\label{CP}
\end{eqnarray}
When $|\Phi| <\Phi_*=\sqrt{3}\pi/{8}$, we have $T_\text{c}>0$ and $P_\text{c}>0$. While when $|\Phi| \ge \Phi_*$, the $T_\text{c}$ and $P_\text{c}$ will be negative and thus the critical point will be unphysical. In section \uppercase\expandafter{\romannumeral4}, we will see that the value of $\Phi_*$ not only affects the phase transition behavior of the black hole, but also changes the interaction force between the black hole molecules.

\begin{figure}[h]
	 \center{\subfigure[]{\label{pvdiagram}\includegraphics[width=7cm,height=5cm]{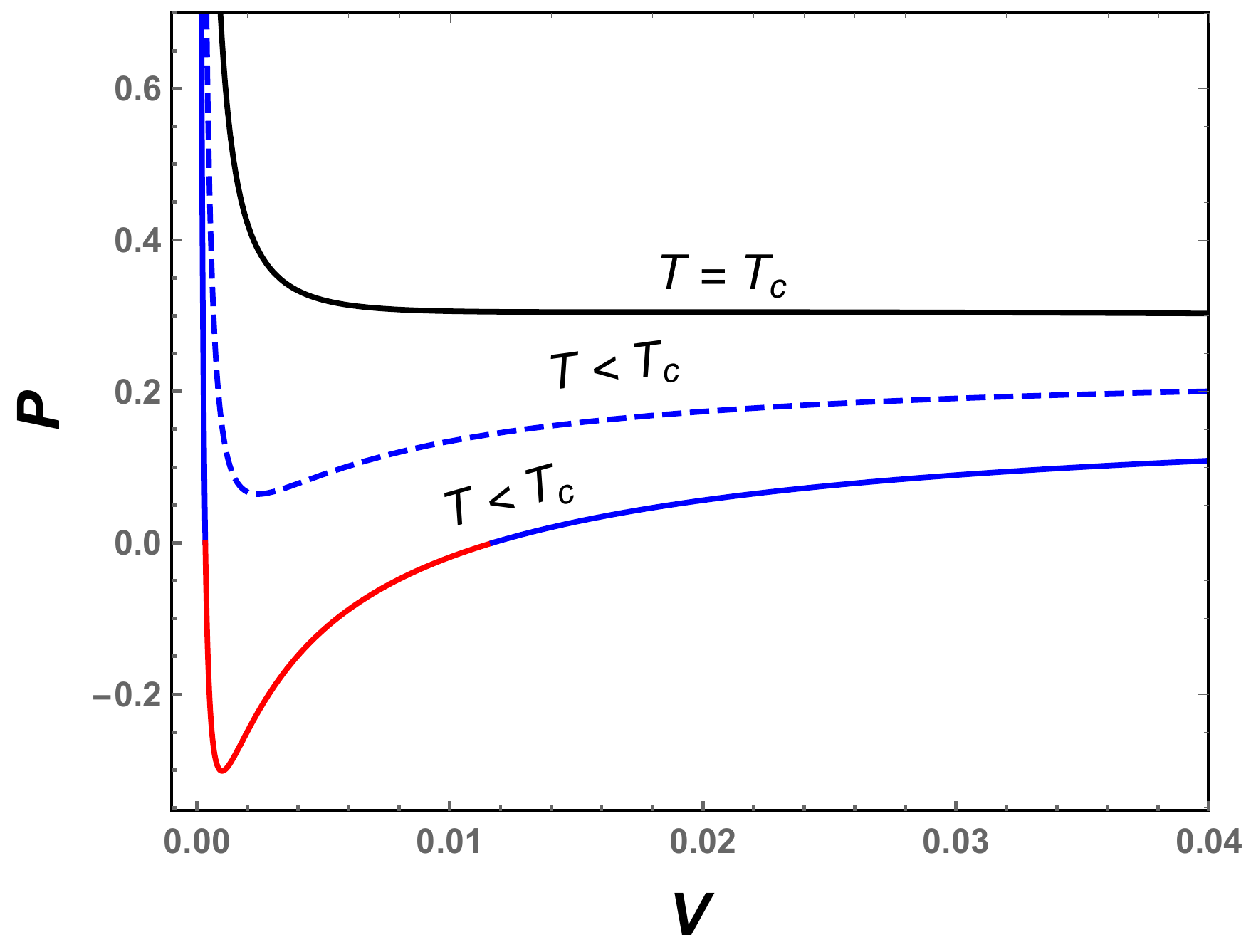}}
		 \subfigure[]{\label{statecurve}\includegraphics[width=7cm,height=5cm]{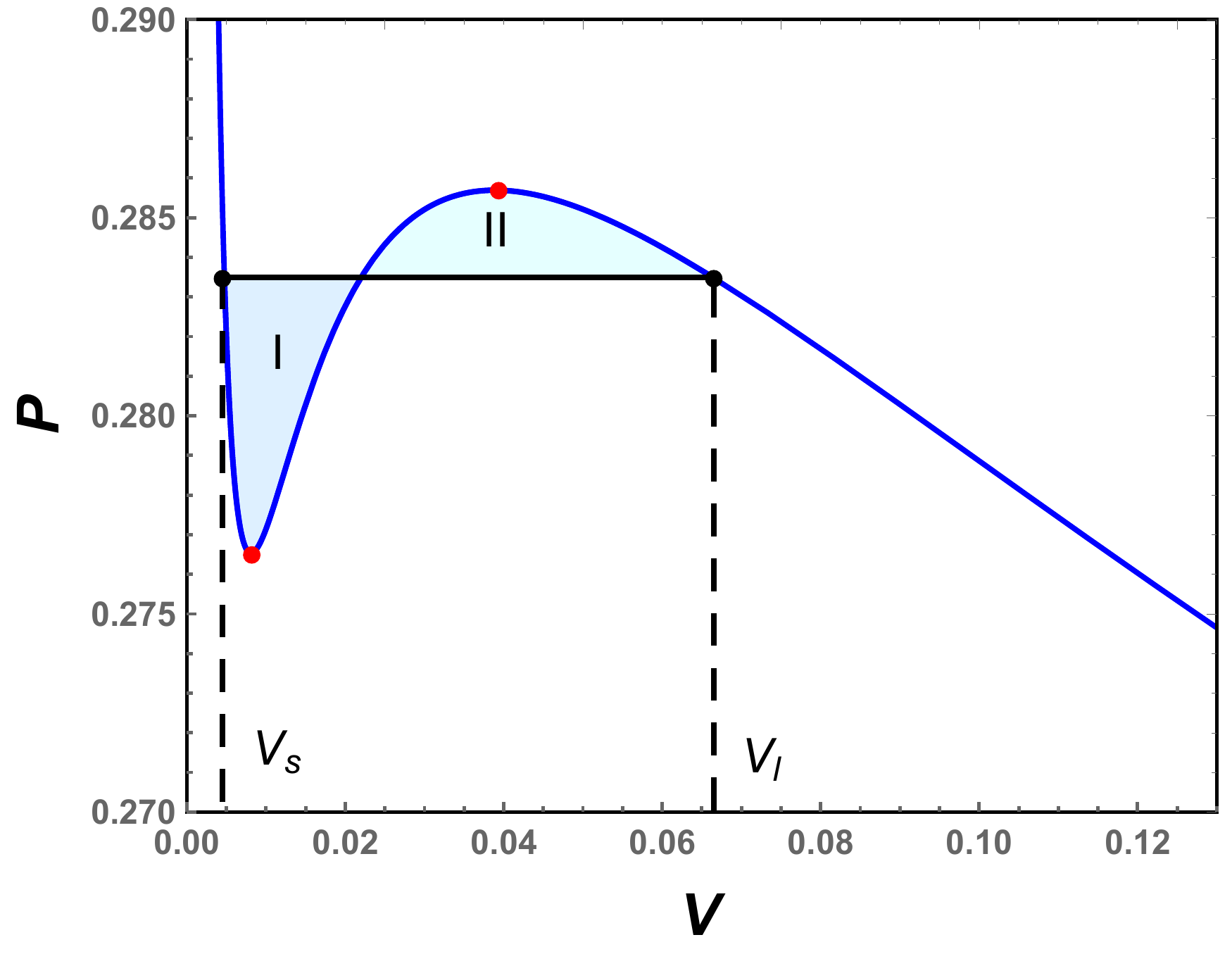}}}
\caption{(a) The isotherms in the $P$-$V$ plane with different values of the temperature $T$. The red part of the curve corresponds to negative pressure. (b) The equal area law in the $P$-$V$ plane at temperature $T$ = 0.293, where areas I and II are equal. $V_s$ and $V_l$ correspond to the volumes of the coexistence small and large black holes, respectively. The electric potential $\Phi$ = 0.5 and GB coupling $\alpha$ = 0.01.} \label{figure}
\end{figure}

\section{Equal area law and phase diagram}
\label{phase}

In this section, we would like to construct the equal area law in the $P$-$V$ plane and then obtain the analytical coexistence curve in the grand canonical ensemble, where the electric potential $\Phi$ is fixed.

There are two reasons why we need the equal area law.

\begin{itemize}
\item There may be a negative pressure part of the isothermal curve when $T < T_\text{c}$ (see Fig. \ref{pvdiagram}). We believe that negative pressure is unphysical and needs to be removed.

\item When the temperature is lower than the critical temperature, there is an unstable region ($\left (\partial_{V} P\right )_{T, \alpha, \Phi} > 0$) between the inflection points of the isothermal curve (see Fig. \ref{statecurve}).
\end{itemize}
As done by Maxwell, we can draw an appropriate horizontal line for each isothermal curve. It is required that these two areas constructed by the curve and horizontal line are equal. Then the temperature and pressure corresponding to the horizontal line is that of the phase transition. This technique is called the Maxwell equal area law. Adopting this law, the above two problems will be naturally solved.

\begin{figure}[h]
	\center{\includegraphics[width=7cm,height=5cm]{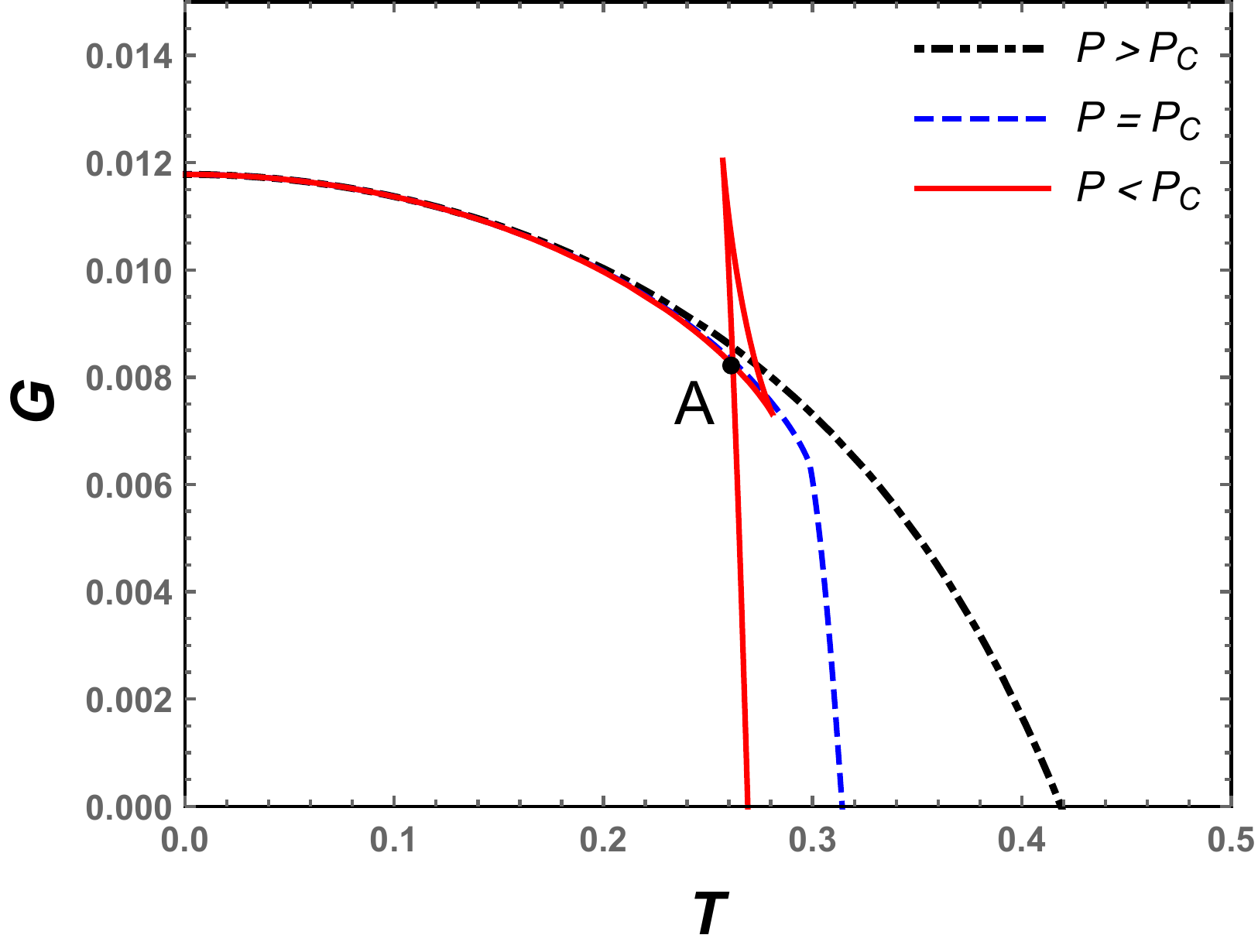}
 \caption{The Gibbs free energy for $P>P_\text{c}$ (black dot-dashed line), $P$ = $P_\text{c}$ (blue dashed line) and $P<P_\text{c}$ (red solid line), where $A$ is an intersection point}.\label{GT}}	    	
\end{figure}

Alternatively, the phase transition can be obtained by examining the behavior of the Gibbs free energy. In the grand canonical ensemble, the Gibbs free energy is defined as
\begin{equation}
G=H-TS-\Phi Q.\label{gibbs}
\end{equation}
We depict $G$ in Fig. \ref{GT} as a function of the temperature. When the temperature is lower than the critical temperature, there presents a swallowtail behavior indicating a first-order phase transition. While for higher temperature the swallowtail behavior disappears. Therefore, by determining the intersection point $A$ of the swallowtail behavior, we can obtain the temperature and pressure of the phase transition. As shown in Ref. \cite{Wei2015c}, the phase transition point determined by the equal area law and the swallowtail behavior of the Gibbs free energy are consistent with each other. It is worth noting that here we exclude the pure thermal radiation phase in our study.

\subsection{Coexistence curve and equal area law}

Here we construct the equal area law in the $P$-$V$ plane and obtain the analytical coexistence curve in the grand canonical ensemble.

\begin{figure}[t]
\center{\subfigure[]{\label{ptcoexistence}\includegraphics[width=7cm,height=5cm]{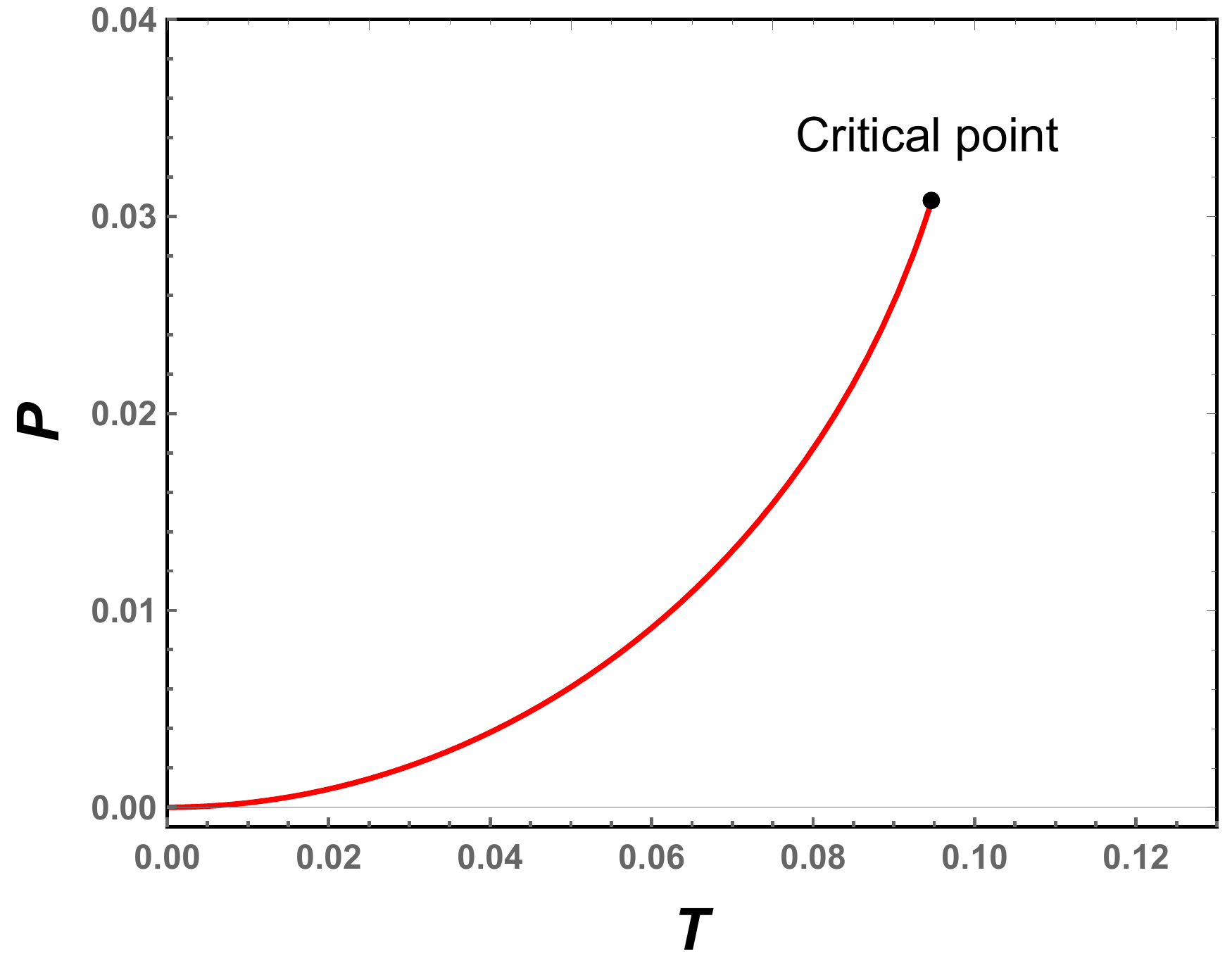}}
\subfigure[]{\label{pvcoexistence}\includegraphics[width=7cm,height=5cm]{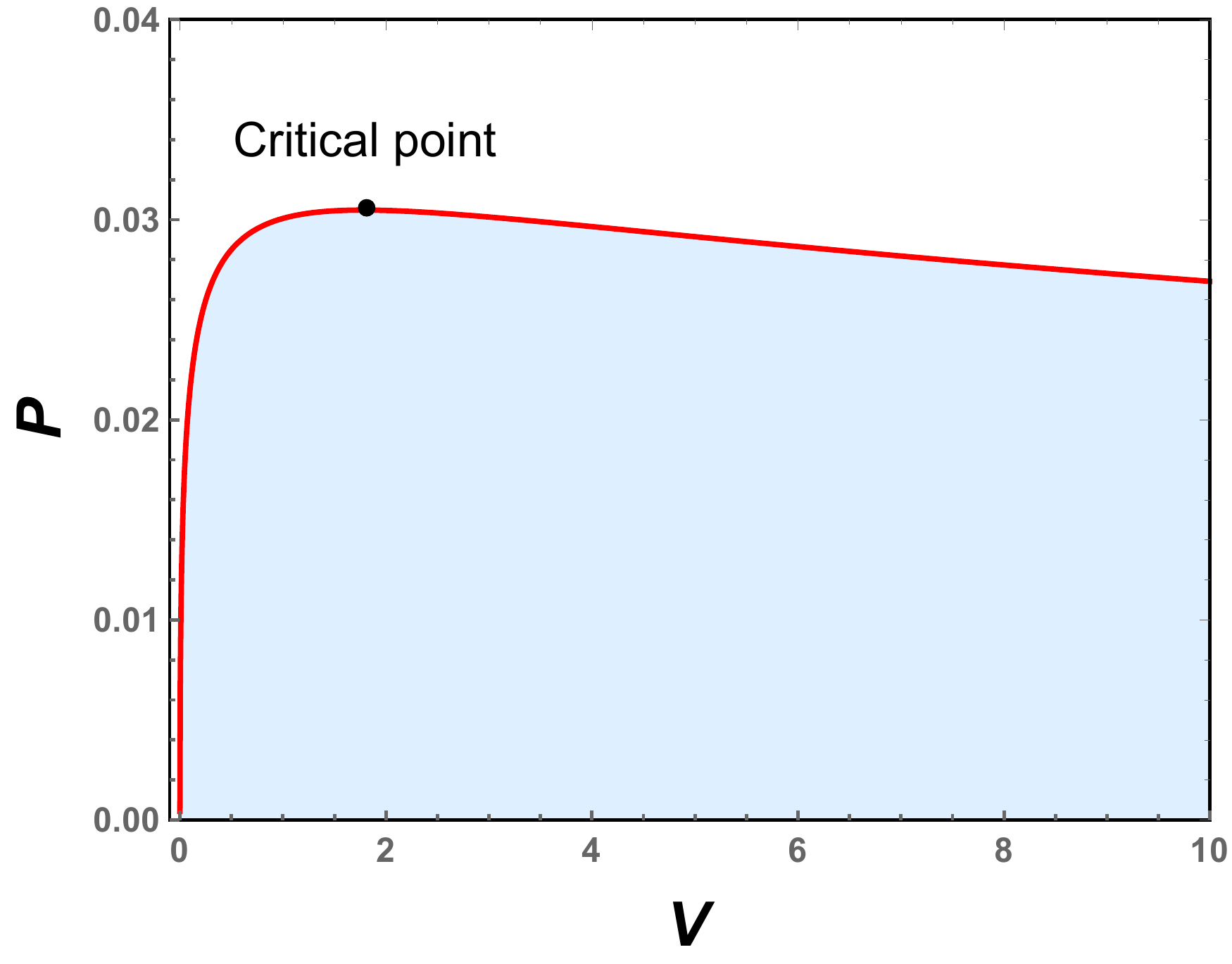}}}
	\caption{Phase diagram for the charged GB-AdS black hole in the grand canonical ensemble with  $\Phi$=0.5 and $\alpha$=0.1. (a) The coexistence curve in the $P$-$T$ plane. (b) The coexistence curve in the $P$-$V$ plane. The shadow region is the coexistence phase of the small and large black holes.  Black dots denote the critical point.}
	\label{co-curve}
\end{figure}

By using the first law of the black hole, the Gibbs free energy has the following differential form
\begin{eqnarray}
dG=-S dT+V dP+\mathcal{A} d\alpha-Qd\Phi.
\end{eqnarray}
Considering that $\text{E}$ and $\text{E}^\prime$ are two thermodynamic coexistence states of a first order phase transition, one easily has $\Delta G=G_{\text{E}}-G_{\text{E}'}=0$. Integrating from state $\text{E}$ to state $\text{E}^\prime$, we have
\begin{align}
-\int_{T_\text{E}}^{T_{\text{E}^\prime}}S dT+\int_{P_\text{E}}^{P_{\text{E}^\prime}}V dP+\int_{\alpha_\text{E}}^{\alpha_{\text{E}^\prime}}\mathcal{A} d\alpha-\int_{\Phi_\text{E}}^{\Phi_{\text{E}^\prime}}Q d\Phi=\int_{G_\text{E}}^{G_{\text{E}^\prime}} dG=0.
\end{align}
At a fixed $T$, $\alpha$, and $\Phi$, one can obtain the following equal area condition
\begin{eqnarray}
\int_{P_\text{E}}^{P_{\text{E}^\prime}}V dP=0,
\end{eqnarray}
or,
\begin{eqnarray}
\int_{V_\text{E}}^{V_{\text{E}^\prime}}P dV=P_{V_\text{E}} (V_{\text{E}^\prime}-V_\text{E}),\label{equal-law}
\end{eqnarray}
where $P_{V_\text{E}}$ is the pressure of the phase transition. Here the black hole admits a small-large black hole phase transition, so we mark $V_\text{E}$ as $V_\text{s}$ and $V_{\text{E}^\prime}$ as $V_\text{l}$ for simplicity. Further we denote the phase transition pressure $P_{V_\text{E}}$ as $P^*$. Eq. (\ref{equal-law}) actually describes the Maxwell equal area law. For example these two areas I and II depicted in Fig. \ref{statecurve} are equal. Inserting Eq. (\ref{statep}) into Eq. (\ref{equal-law}), we have
\begin{eqnarray}
P^*(V_\text{l}-V_\text{s})=3 \times 2^{1 / 4} \pi^{3 / 2}T\left(V_\text{l}^{1 / 4}-V_\text{s}^{1 / 4}\right) \alpha+\frac{\left(-3 \pi^{2}+64 \Phi^{2}\right)}{4 \sqrt{2} \pi^{2}}(V_\text{l}^{1/2}-V_\text{s}^{1/2})+\frac{\sqrt{\pi} }{2^{1 / 4}}T\left(V_\text{l}^{3 / 4}-V_\text{s}^{3 / 4}\right).\label{p1}
\end{eqnarray}
Moreover, the small and large black hole states satisfy the state equation, which gives
\begin{eqnarray}
P^*&=&\frac{\left(64 \Phi^{2}-3 \pi^{2}\right) \sqrt{2}}{16 \pi^{2}V_\text{s}^{\frac{1}{2}}} +\frac{3 \times 2^{\frac{3}{4}} \pi^{\frac{1}{2}} T}{8V_\text{s}^{\frac{1}{4}}} +\frac{3 \times 2^{\frac{1}{4}} \pi^{\frac{3}{2}} T \alpha}{4V_\text{s}^{\frac{3}{4}}} ,\label{p2}\\
P^*&=&\frac{\left(64 \Phi^{2}-3 \pi^{2}\right) \sqrt{2}}{16 \pi^{2}V_\text{l}^{\frac{1}{2}}} +\frac{3 \times 2^{\frac{3}{4}} \pi^{\frac{1}{2}} T}{8V_\text{l}^{\frac{1}{4}}} +\frac{3 \times 2^{\frac{1}{4}} \pi^{\frac{3}{2}} T \alpha}{4V_\text{l}^{\frac{3}{4}}} .\label{p3}
\end{eqnarray}
Solving these three equations (\ref{p1}), (\ref{p2}) and (\ref{p3}), we will obtain the coexistence curve of the small and large black holes. In order to solve these equations, we denote $V_\text{s}=a^4$ and $V_\text{l}=b^4$. Then the Eqs.(\ref{p1})-(\ref{p3}) reduce to
\begin{eqnarray}
P^*&=&\frac{1}{-a^{4}+\mathrm{b}^{4}}\left(-\frac{\left(a^{3}-\mathrm{b}^{3}\right) \sqrt{\pi} }{2^{1 / 4}}\mathrm{T}+3 \times 2^{1 / 4}(-a+\mathrm{b}) \pi^{3 / 2} \mathrm{T} \alpha+\frac{\left(a^{2}-\mathrm{b}^{2}\right)\left(3 \pi^{2}-64 \Phi^{2}\right)}{4 \sqrt{2} \pi^{2}}\right),\label{p4}\\
P^*&=&\frac{6 \times 2^{3 / 4} a^{2} \pi^{5 / 2} \mathrm{T}+12 \times 2^{1 / 4} \pi^{7 / 2} \mathrm{T} \alpha+\sqrt{2} a\left(-3 \pi^{2}+64 \Phi^{2}\right)}{16 a^{3} \pi^{2}},\label{p5}\\
P^*&=&\frac{6 \times 2^{3 / 4} \mathrm{b}^{2} \pi^{5 / 2} \mathrm{T}+12 \times 2^{1 / 4} \pi^{7 / 2} \mathrm{T} \alpha+\sqrt{2} \mathrm{b}\left(-3 \pi^{2}+64 \Phi^{2}\right)}{16 \mathrm{b}^{3} \pi^{2}}.\label{p6}
\end{eqnarray}
Combining with them, we arrive at
\begin{eqnarray}\label{ab1}
&&12\left(a^{4}+3 a^{3} b+6 a^{2} b^{2}+3 a b^{3}+b^{4}\right)\pi^{7/2}T\alpha \nonumber\\
&&+2^{1 / 4} a b\left(-3(a+b)^{3} \pi^{2}+2\times2^{1 / 4} a b\left(3 a^{2}+4 a b+3 b^{2}\right) \pi^{5 / 2} T+64(a+b)^{3} \Phi^{2}\right)=0,\\
&&12\left(a^{3}-b^{3}\right) \pi^{7 / 2} T \alpha+2^{1 / 4} a(a-b) b\left(-3(a+b) \pi^{2}+6\times2^{1 / 4} a b \pi^{5 / 2} T+64(a+b) \Phi^{2}\right)=0.\label{ab2}
\end{eqnarray}
Further, Eqs. (\ref{ab1}) and (\ref{ab2}) can be expressed as
\begin{eqnarray}
T&=& \frac{a b(a+b)^{3}\left(3 \pi^{2}-64 \Phi^{2}\right)}{2 \times 2^{1 / 4} \pi^{5 / 2}\left(a^{2} b^{2}\left(3 a^{2}+4 a b+3 b^{2}\right)+3 \sqrt{2}\left(a^{4}+3 a^{3} b+6 a^{2} b^{2}+3 a b^{3}+b^{4}\right) \pi \alpha\right)},\label{T1}\\
T&=&\frac{a b(a+b)\left(3 \pi^{2}-64 \Phi^{2}\right)}{6 \times 2^{1 / 4} \pi^{5 / 2}\left(a^{2} b^{2}+\sqrt{2}\left(a^{2}+a b+b^{2}\right) \pi \alpha\right)}.\label{T2}
\end{eqnarray}
Combing Eqs. (\ref{T1}) and (\ref{T2}), we have the relation
\begin{equation}
a b=3 \sqrt{2} \pi \alpha.
\end{equation}
By substituting this relation into Eq. (\ref{ab1}), we get
\begin{eqnarray}
V_\text{s}&=&\left(\frac{3 \pi^{2}-64 \Phi^{2}+X-T \sqrt{-384 \pi^{6} \alpha+\frac{(3 \pi^{2}-64 \Phi^{2}+X)^2}{T^{2}}}}{8 \times 2^{1 / 4} \pi^{5 / 2} T}\right)^4,\label{vs}\\
V_\text{l}&=&\left(\frac{3 \pi^{2}-64 \Phi^{2}+X+T \sqrt{-384 \pi^{6} \alpha+\frac{(3 \pi^{2}-64 \Phi^{2}+X)^2}{T^{2}}}}{8 \times 2^{1 / 4} \pi^{5 / 2} T}\right)^4,\label{vl}
\end{eqnarray}
where
\begin{eqnarray}
X=\sqrt{9 \pi^{4}-192 \pi^{6} T^{2} \alpha-384 \pi^{2} \Phi^{2}+4096\Phi^{4}}.
\end{eqnarray}
Plugging $V_\text{s}$ and $V_\text{l}$ into Eq. (\ref{p1}), we obtain the analytical form of the coexistence curve in the $P$-$T$ plane
\begin{eqnarray}
P=-\frac{-3 \pi^{2}+64 \Phi^{2}+\sqrt{9 \pi^{4}-192 \pi^{6} T^{2} \alpha-384 \pi^{2} \Phi^{2}+4096 \Phi^{4}}}{96 \pi^{3} \alpha}.\label{CC1}
\end{eqnarray}
The coexistence curve is described in Fig. \ref{ptcoexistence}. Above the curve is the small black hole phase and below it is for the large black hole phase. The curve starts at the origin and ends at the critical point denoted by a black dot. Moreover, in the $P$-$V$ and $T$-$V$ planes, the coexistence curves read
\begin{eqnarray}
P&=&\frac{\left(\sqrt{2} V^{3 / 2}-24 \pi V \alpha+18 \sqrt{2} \pi^{2} V^{1/2} \alpha^{2}\right)\left(3 \pi^{2}-64 \Phi^{2}\right)}{8 \pi^{2}\left(V^{2}-252 \pi^{2} V \alpha^{2}+324 \pi^{4} \alpha^{4}\right)},\label{CC2}\\
T &=& \frac{ V^{1 / 4}\left(\sqrt{2} V^{3 / 2}-18 \pi V\alpha-54 \sqrt{2} \pi^{2} \sqrt{ V} \alpha^{2}+108 \pi^{3} \alpha^{3}\right)\left(3 \pi^{2}-64 \Phi^{2}\right)}{2 \times 2^{3 / 4} \pi^{5 / 2}\left( V^{2}-252 \pi^{2} V \alpha^{2}+324 \pi^{4} \alpha^{4}\right)}.\label{CTV}
\end{eqnarray}
For an example, we show the phase diagram in the $P$-$V$ plane in Fig. \ref{pvcoexistence}. The shadow region denotes the coexistence regions of the small and large black holes. The left and right regions are for the small and large black holes, respectively.  In the $T$-$V$ plane, the phase diagram has similar shape.

\subsection{Critical exponent}

As we know, the critical exponents reveal the universal properties of the system near the critical point. So in this section, we would like to examine them.

From Eqs. (\ref{vs}) and (\ref{vl}), it is obvious that $V_\text{s}$ and $V_\text{l}$ have analytical forms. Expanding them around the critical point, we have
\begin{eqnarray}
	(V_\text{s}-V_\text{c})&=&-\frac{432\left(6^{1 / 4} \pi^{7 / 2} \alpha^{9 / 4}\right)}{\sqrt{3 \pi^{2}-64 \Phi^{2}}}(T_{c}-T)^{1/2}+\frac{5184 \sqrt{6} \pi^{5} \alpha^{5 / 2} }{3 \pi^{2}-64 \Phi^{2}}(T_{c}-T)+\mathcal{O}(T_{c}-T)^{3 / 2},\label{vt1}\\
   (V_\text{l}-V_\text{c})&=&\frac{432\left(6^{1 / 4} \pi^{7 / 2} \alpha^{9 / 4}\right)}{\sqrt{3 \pi^{2}-64 \Phi^{2}}}(T_{c}-T)^{1/2}+\frac{5184 \sqrt{6} \pi^{5} \alpha^{5 / 2} }{3 \pi^{2}-64 \Phi^{2}}(T_{c}-T)+\mathcal{O}(T_{c}-T)^{3 / 2}.\label{vt2}
\end{eqnarray}
Obviously, near the critical point, $(V_\text{s}-V_\text{c})$ and $(V_\text{l}-V_\text{c})$ share the same critical exponent of $\frac{1}{2}$. These coefficients depend on the electric potential $\Phi$ and GB coupling $\alpha$. Interestingly, the absolute values of these two coefficients are equal to each other.

\begin{figure}[t]
\center{\subfigure[]{\label{ordervt}\includegraphics[width=7cm,height=5cm]{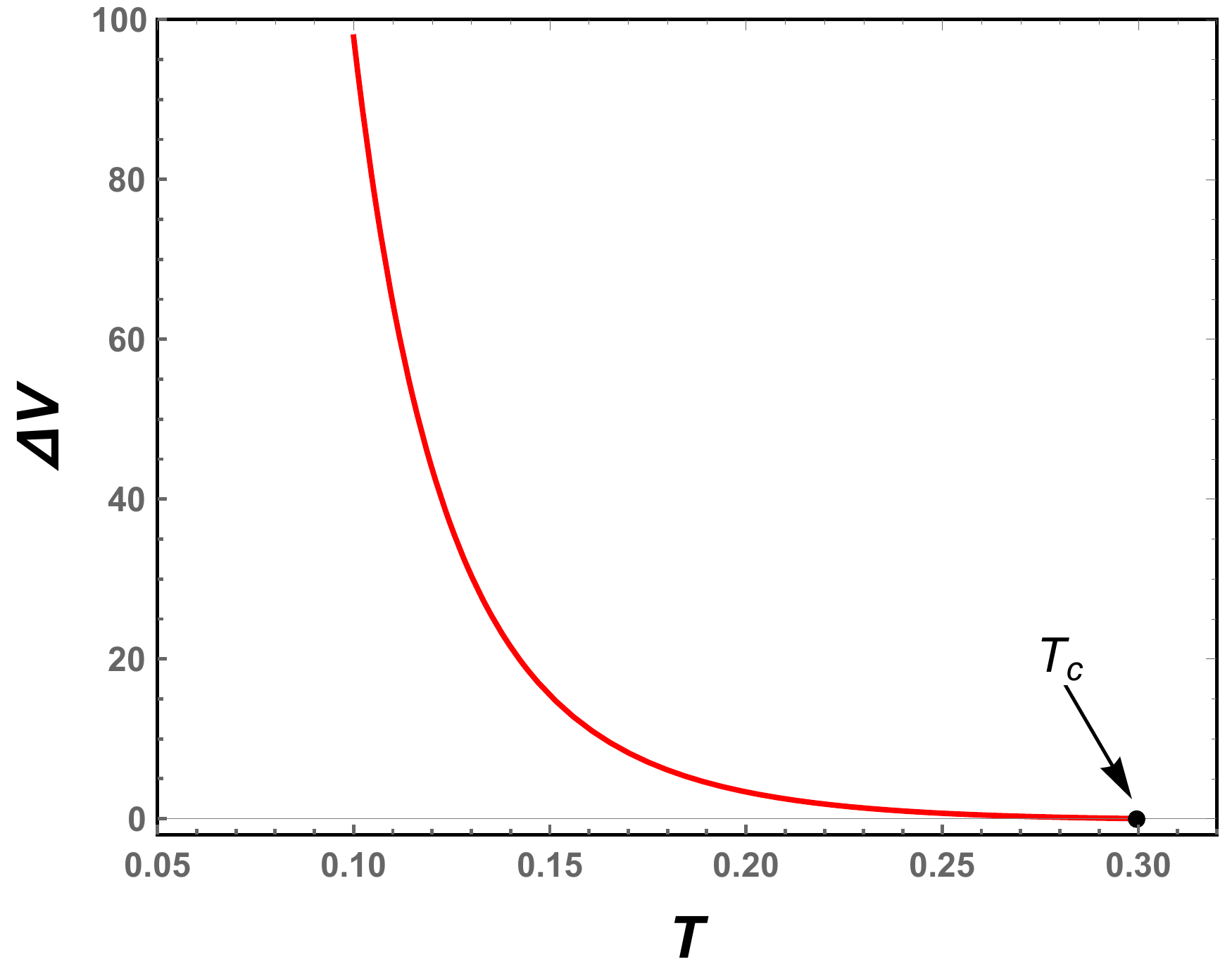}}
 \subfigure[]{\label{ordervp}\includegraphics[width=7cm,height=5cm]{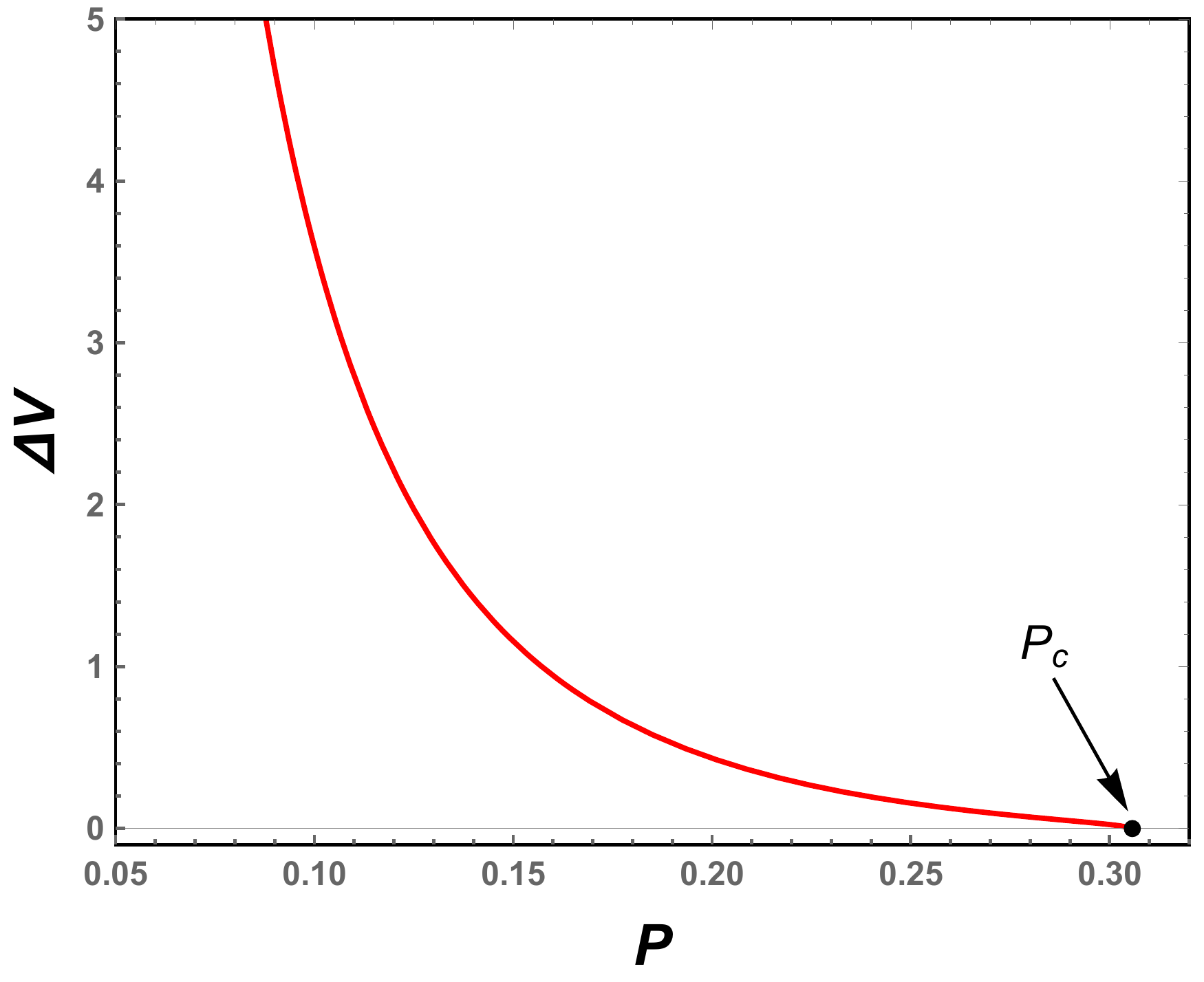}}}
 \caption{The change of the thermodynamic volume $\Delta V$ among the black hole phase transition with $\Phi$=0.5 and $\alpha$=0.01. \subref{ordervt} $\Delta V$ vs. $T$. \subref{ordervp} $\Delta V$ vs. $P$.}
	\label{order}
\end{figure}

On the other hand, we introduce $\Delta V = V_\text{l}-V_\text{s}$ to denote the change of the volume among the phase transition. Its behavior is shown in Fig. \ref{order}. We observe that $\Delta V$ decreases with the temperature or the pressure. While when the critical point is approached, $\Delta V$=0, indicating that the small and large black hole phases cannot be clearly distinguished. Combining with Eqs. (\ref{vt1}) and  (\ref{vt2}), we obtain
\begin{eqnarray}
 \Delta V=\frac{864 \times 6^{1 / 4} \pi^{7 / 2} \alpha^{9 / 4}}{\sqrt{3 \pi^{2}-64 \Phi^{2}}}(T_{c}-T)^{1 / 2}+\mathcal{O}(T_{c}-T)^{3 / 2}.\label{VTE}
\end{eqnarray}
This reveals that when $T=T_c$, $\Delta V$=0. Moreover, at the critical point, $\Delta V$ has a critical exponent $\frac{1}{2}$.

Actually, the critical exponent can also be obtained around the critical pressure. Solving Eq. (\ref{CC2}), one easily gets
\begin{eqnarray}
V_\text{s}&=&\frac{9 \pi^{4}(1+64 P \pi \alpha(14 P \pi \alpha-1))+384 \pi^{2}(32 P \pi \alpha-1) \Phi^{2}+4096 \Phi^{4}-Y}{64 P^{2} \pi^{4}},\label{vps}\\
V_\text{l}&=&\frac{9 \pi^{4}(1+64 P \pi \alpha(14 P \pi \alpha-1))+384 \pi^{2}(32 P \pi \alpha-1) \Phi^{2}+4096 \Phi^{4}+Y}{64 P^{2} \pi^{4}},\label{vpl}
\end{eqnarray}
where
\begin{eqnarray}
Y=\sqrt{\left(-3 \pi^{2}+48 P \pi^{3} \alpha+64 \Phi^{2}\right)\left(-3 \pi^{2}+96 P \pi^{3} \alpha+64 \Phi^{2}\right)^{2}\left(3 \pi^{2}(48 P \pi \alpha-1)+64 \Phi^{2}\right)}.
\end{eqnarray}
Expanding them around the critical pressure $P_c$, we obtain
\begin{eqnarray}
(V_\text{s}-V_\text{c})&=&-\frac{432\left(\sqrt{6} \pi^{7 / 2} \alpha^{5 / 2}\right) }{\sqrt{3 \pi^{2}-64 \Phi^{2}}}(P_\text{c}-P)^{1/2}+\frac{31104 \pi^{5} \alpha^{3} }{3 \pi^{2}-64 \Phi^{2}}(P_\text{c}-P)+\mathcal{O}(P_\text{c}-P)^{3 / 2},\label{CV1}\\
(V_\text{l}-V_\text{c})&=&\frac{432\left(\sqrt{6} \pi^{7 / 2} \alpha^{5 / 2}\right) }{\sqrt{3 \pi^{2}-64 \Phi^{2}}}(P_\text{c}-P)^{1/2}+\frac{31104 \pi^{5} \alpha^{3} }{3 \pi^{2}-64 \Phi^{2}}(P_\text{c}-P)+\mathcal{O}(P_\text{c}-P)^{3 / 2}.\label{CV2}
\end{eqnarray}
Obviously, the critical exponent is $\frac{1}{2}$, which is the same as that around the critical temperature. Similarly, near the critical pressure, $\Delta V$ has the following form
\begin{eqnarray}
\Delta V=\frac{864 \sqrt{6} \pi^{7 / 2} \alpha^{5 / 2} }{\sqrt{3 \pi^{2}-64 \Phi^{2}}}(P_\text{c}-P)^{1/2}+\mathcal{O}(P_\text{c}-P)^{3 / 2}.\label{VPE}
\end{eqnarray}
The critical exponent keeps unchanged. The detailed behavior of $\Delta V$ can also be found in Fig. \ref{ordervp}.

Furthermore, in the reduced parameter space, Eqs. (\ref{VTE}) and (\ref{VPE}) can be simplified to
\begin{eqnarray}
	\Delta \tilde{V}&=&8 \sqrt{6}(1-\tilde{T})^{\frac{1}{2}}+ \mathcal{O} (1-\tilde{T})^{\frac{3}{2}}, \\
	\Delta \tilde{V}&=&4 \sqrt{6}(1-\tilde{P})^{\frac{1}{2}}+ \mathcal{O} (1-\tilde{P})^{\frac{3}{2}},
\end{eqnarray}
where $\Delta \tilde{V}=\frac{\Delta V}{V_c}$, $\tilde{T}=\frac{T}{T_c}$, $\tilde{P}=\frac{P}{P_c}$. This result is identical with the five-dimensional neutral GB-AdS black hole \cite{Wei2020a}, which means that whether the black hole is charged or not, they all behave exactly the same near the critical point in the reduced parameter space.

\section{Ruppeiner geometry}
\label{Ruup}

Although we do not know how quantum gravity theory describes the microscopic states of black holes, we can explore the interaction between black hole molecules by making use of the popular thermodynamic tool---Ruppiner geometry.

The Ruppeiner geometry was introduced to describe interparticle interactions in a thermodynamic system \cite{Ruppeiner1979,Ruppeiner1995}. It was the first to systematically calculate the thermodynamic scalar curvature $R$ \cite{Ruppeiner2014b}. The sign of $R$ corresponds to the interactions between two interparticles of the system. For example, positive or negative $R$ indicates a repulsive or attractive interaction \cite{Ruppeiner2010}, and $R=0$ shows a system without interaction. Moreover, $R$ is also linked to the correlation length near the critical point.

Let us start with the probability  expression of a system fluctuating deviation from equilibrium \cite{Landau1977}
\begin{eqnarray}\label{pb}
&& P_{\rm probability} \propto e^{-\frac{1}{2} \Delta l^{2}},
\end{eqnarray}
with
\begin{eqnarray}
\Delta l^{2}&=&- g_{\mu \nu} \Delta x^{\mu} \Delta x^{\nu},\\
 g_{\mu \nu}&=&\frac{\partial^{2} S}{\partial x^{\mu} \partial x^{\nu}}.\label{metric}
\end{eqnarray}
Here $\Delta l^{2}$ is the thermodynamic line element, $g_{\mu \nu}$ is the thermodynamic metric tensor, $x^{\mu}$ denotes the independent fluctuating thermodynamic variables. As we can see in Eq. (\ref{pb}), the smaller the line element $\Delta l^{2}$ is, the greater the probability of a fluctuation away from equilibrium, and thus it means that the line element $\Delta l^{2}$ measures the distance between two neighboring fluctuation states in the thermodynamic parameter space.

Since (\ref{metric}) is similar to the metric tensor in the Riemannian geometry, it allows us to construct the Christoffel symbol, Riemann curvature tensor, Ricci tensor, and Riemann scalar curvature, which are given by
\begin{eqnarray}
&& \Gamma_{\mu \nu}^{\sigma}\;\;= \frac{1}{2} g^{\sigma \rho}\left(\partial_\nu g_{\rho \mu}+\partial_\mu g_{\nu \rho}-\partial_\rho g_{\mu \nu}\right),\label{CS}\\
&& R_{\mu \nu \sigma}^{\rho}= \partial_\mu\Gamma_{\nu \sigma}^{\rho}-\partial_\nu\Gamma_{\mu \sigma}^{\rho}+\Gamma_{\sigma \nu}^{\lambda} \Gamma_{\mu \lambda}^{\rho}-\Gamma_{\sigma \mu}^{\lambda} \Gamma_{\nu \lambda}^{\rho},\label{RCT}\\
&& R_{\mu \nu}\;\;= R_{\mu \lambda \nu}^{\lambda}, \label{RT}\\
&& R\quad\;\;= g^{\mu \nu} R_{\mu \nu}. \label{RCS}
\end{eqnarray}
Note that in Eq. (\ref{RCT}), we can also define a Riemann curvature tensor with the opposite sign. In this paper, we adopt the same definition as that given in \cite{Ruppeiner1995}, where $R$ is positive (negative) when the interaction between particles is repulsive (attractive).

Now we choose temperature $T$ and thermodynamic volume $V$ as the fluctuation coordinates and thus $x^1=T$, $x^2=V$. Then the line element can be expressed in the following form \cite{Wei2019d, Wei2020a}
\begin{eqnarray}
\Delta l^{2}=-\frac{1}{T}\left(\frac{\partial^{2} F}{\partial T^{2}}\right) \Delta T^{2}+\frac{1}{T}\left(\frac{\partial^{2} F}{\partial V^{2}}\right) \Delta V^{2},
\end{eqnarray}
where the Helmholtz free energy is $F=U-T S-Q \Phi$ with $U$ the internal energy of the system. We also have $d F=-S d T-P d V+\mathcal{A} d \alpha-Q d \Phi$. By using the heat capacity at constant volume, $C_\text{V}=T\left(\partial_{T} S\right)_{V}$, the line element will be of the following form
\begin{eqnarray}
\Delta l^{2}=\frac{C_\text{V}}{T^{2}} \Delta T^{2}-\frac{\left(\partial_{V} P\right)_{T}}{T} \Delta V^{2}.\label{le}
\end{eqnarray}
Then following Eqs. (\ref{CS})-(\ref{RCS}), the scalar curvature can be calculated as
\begin{eqnarray}
R&=&\frac{1}{2 C_\text{V}^{2}(\partial_{V} P)^{2}}\left[T\left(\partial_{V} C_\text{V}\right)^{2} (\partial_{V} P)-T C_\text{V}( \partial_{V} C_\text{V})\left(\partial_{V, V} P\right)^{2}+T (\partial_{V} P)( \partial_{T} C_\text{V})\left(\partial_{V} P-T \partial_{T, V} P\right)\right.\nonumber\\
&+&\left.C_\text{V}\left((\partial_{V} P)^{2}-T^{2}\left(\partial_{T, V} P\right)^{2}-2 T( \partial_{V} P)\left(\partial_{V, V} C_\text{V}-T \partial_{T, T, V} P\right)\right)\right].
\end{eqnarray}
From Eqs. (\ref{entropy}) and (\ref{volume}), we can see that when fixing the volume $V$ and the GB coupling $\alpha$, $dS$ equals to zero, and thus the heat capacity $C_\text{V} = T\left(\frac{\partial S}{\partial T}\right)_{\text{V}}$ vanishes. Under this case, the matric coefficient for the first term of Eq. (\ref{le}) is zero and its inverse is diverging. In order to remove the influence of vanishing $C_\text{V}$ and uncover the black hole microstructure, we follow the treatment of Ref. \cite{Wei2019} and construct the normalized scalar curvature
\begin{eqnarray}
R_\text{N}&=&R*C_\text{V}\nonumber\\
&=&\frac{\left(\partial_{V} P\right)^{2}-T^{2}\left(\partial_{T, V} P\right)^{2}+2 T^2( \partial_{V} P)\left( \partial_{T, T, V} P\right)}{2 \left(\partial_{V} P\right)^{2}}.\label{R-CV}
\end{eqnarray}
Plugging Eq. (\ref{statep}) into it, the normalized scalar curvature becomes
\begin{eqnarray}
R_\text{N}=-\frac{V^{1 / 4}\left(3 \pi^{2}-64 \Phi^{2}\right)\left(-3 \times 2^{1 / 4} \pi^{2} V^{1 / 4}+6 \sqrt{2} \pi^{5 / 2} T \sqrt{V}+36 \pi^{7 / 2} T \alpha+64 \times 2^{1 / 4} V^{1 / 4} \Phi^{2}\right)}{2^{3 / 4}\left(-3 \times 2^{1 / 4} \pi^{2} V^{1 / 4}+3 \sqrt{2} \pi^{5 / 2} T \sqrt{V}+18 \pi^{7 / 2} T \alpha+64 \times 2^{1 / 4} V^{1 / 4} \Phi^{2}\right)^{2}}.\label{RN}
\end{eqnarray}
Combining with the physical interpretation of the scalar curvature, we can test the properties of the black hole microstructure. Next, we will investigate the relationship between the normalized scalar curvature and the electric potential to obtain the information of the black hole microstructure in the grand canonical ensemble. From (\ref{RN}), one can find that $R_\text{N}$ depends on $\Phi^2$. So the properties of the black hole microstructure are only affected by the absolute value of $\Phi$, while ignored with its sign.

\begin{figure}
\center{\includegraphics[width=8cm,height=5cm]{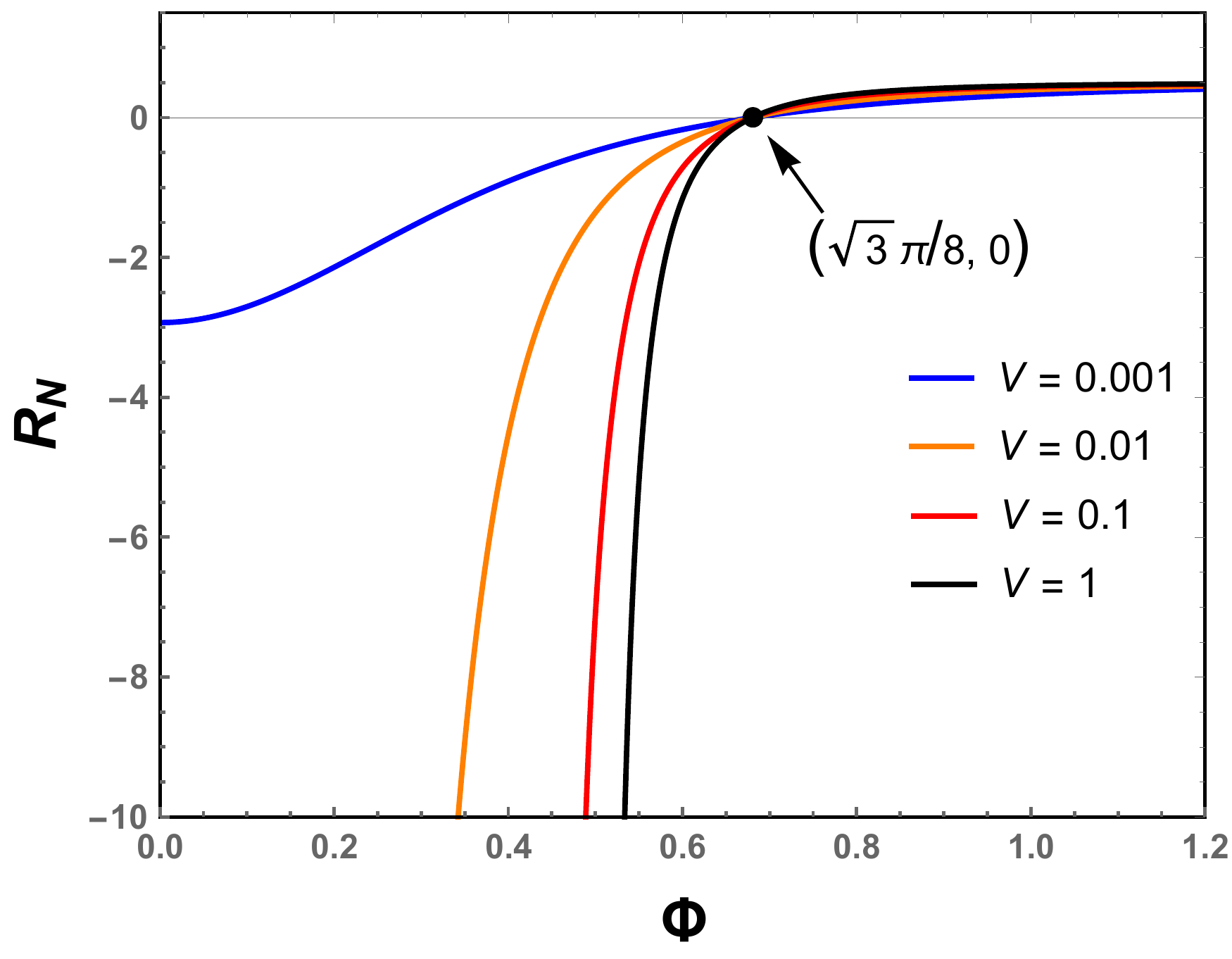}}
	\caption{The normalized scalar curvature $R_\text{N}$ changes with the electric potential $\Phi$ for different volumes. All curves intersect at the same point ($\Phi$, $R_\text{N}$)=($\sqrt{3}\pi/{8}$, 0). We have set $T$ =0.1 and $\alpha$=0.1.} \label{R-Phia}
\end{figure}

Next, we will examine the behavior of the normalized scalar curvature $R_\text{N}$. Let us first consider the case that $R_\text{N}$ changes with the electric potential $\Phi$. For the purpose, we describe the normalized scalar curvature as a function of $\Phi$ for fixed $T$=0.1 and $\alpha$=0.1 in Fig. \ref{R-Phia}. From the figure, we find three interesting phenomena:
\begin{itemize}
\item Each curve of different $V$ merges at the point $(\Phi, R_\text{N})=(\sqrt{3}\pi/{8}, 0)$. Also at this point, $R_\text{N}$ vanishes. This result states that the black hole system at this point is similar to the ideal gas, where no interaction exists among its microstructures. Moreover, from Eq. (\ref{RN}), we find that this property is a universal result and independent of $T$, $V$, and $\alpha$.
\item When $0<\Phi<\sqrt{3}\pi/{8}$, we have $R_\text{N} < 0$, which indicates that in this parameter range, the attractive interaction dominates among the black hole microstructures. It is also can be found that $|R_\text{N}|$ increases with the thermodynamic volume.
\item When $\Phi>\sqrt{3}\pi/{8}$, we observe a positive $R_\text{N}$. Therefore, the dominated interaction is repulsive. $R_\text{N}$ also increases with $V$. One thing worths to note is that in this parameter range the critical temperature and pressure are negative. So the small-large black hole phase transition does not exist.
\end{itemize}

\begin{figure}[h]
	 \center{\subfigure[$\Phi$=0.1]{\label{ctvone}\includegraphics[width=0.45\textwidth]{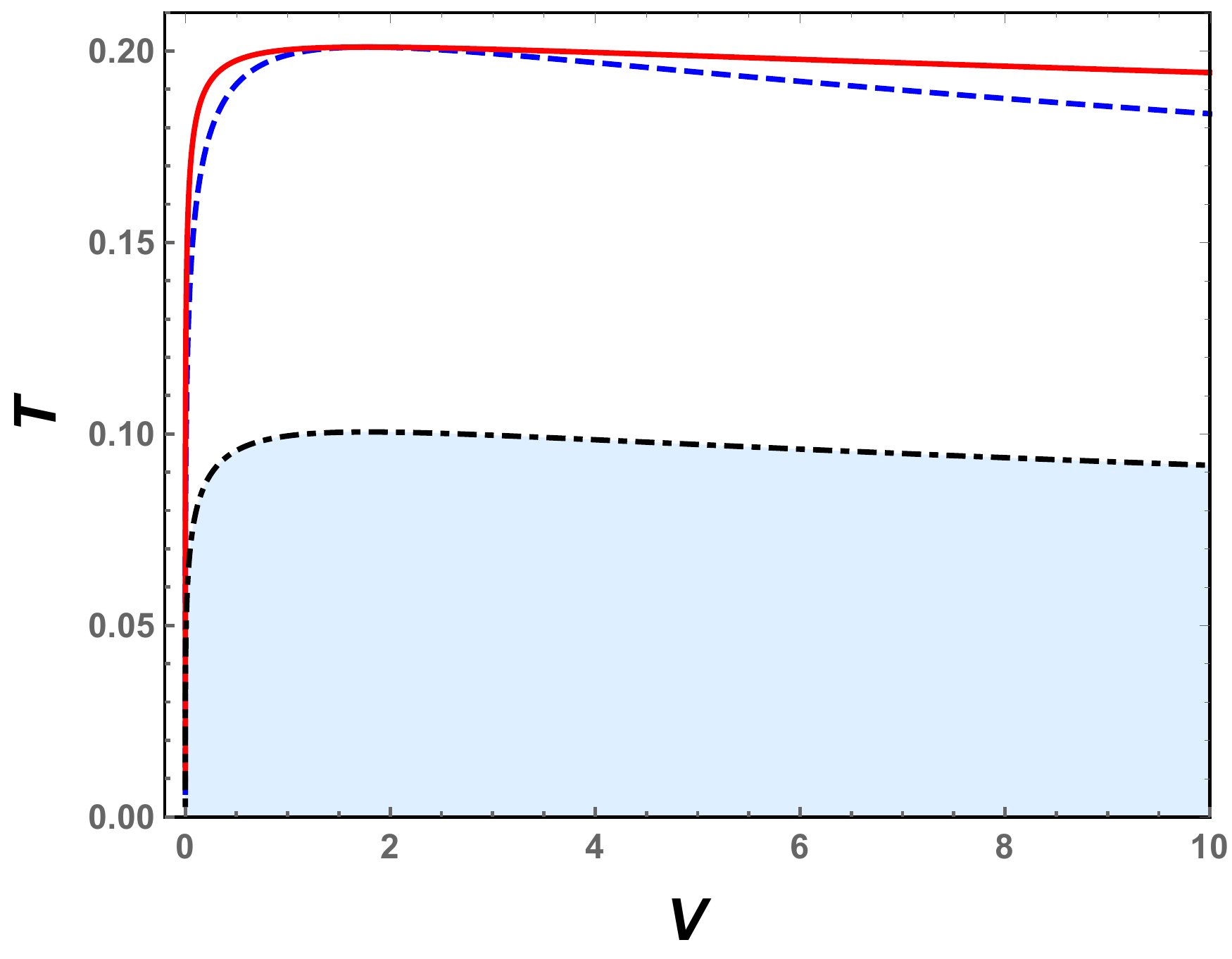}}
		 \subfigure[$\Phi$=0.3]{\label{ctvtwo} \includegraphics[width=0.45\textwidth]{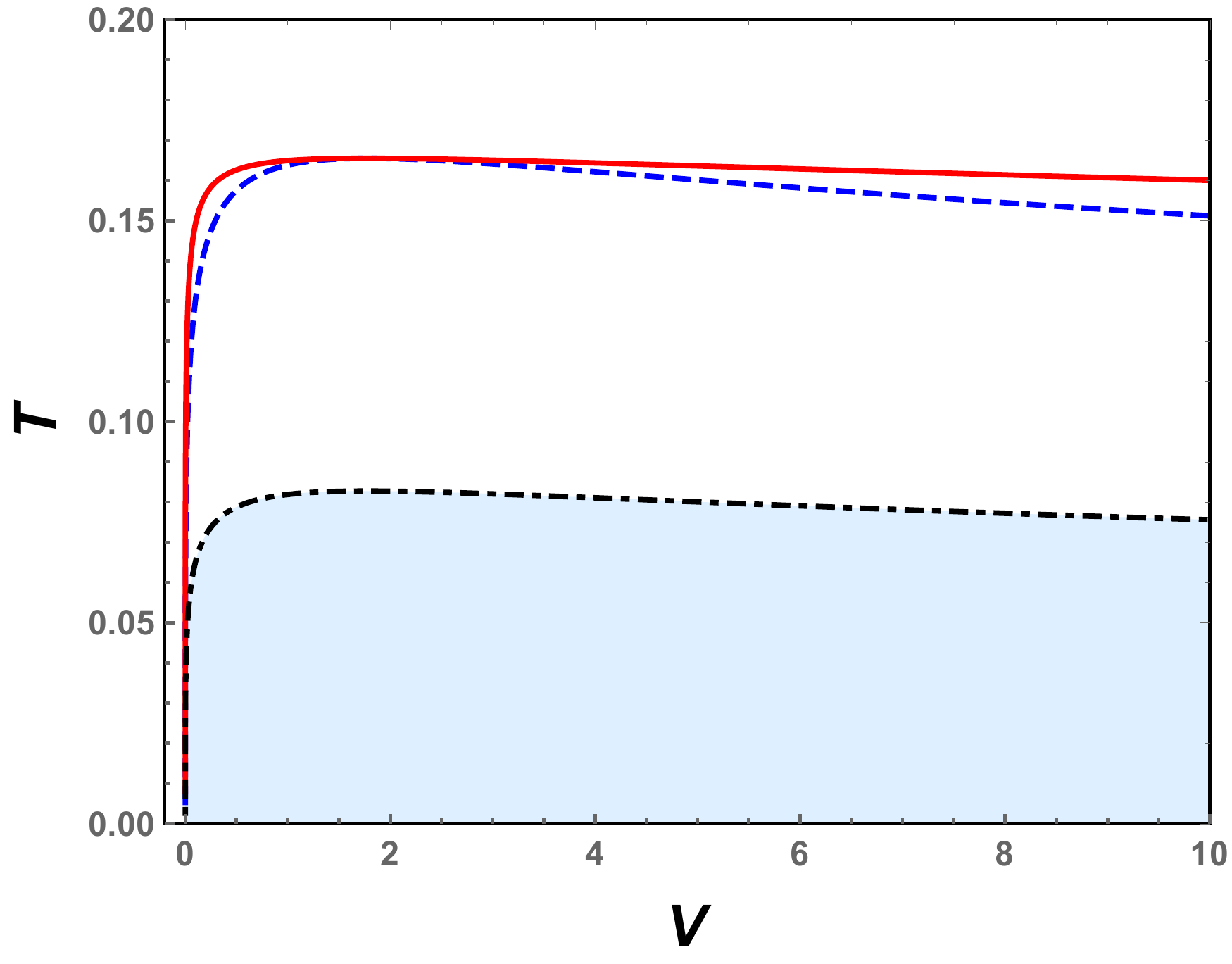}}}\\
	 \subfigure[$\Phi$=0.5]{\label{ctvthree}\includegraphics[width=0.45\textwidth]{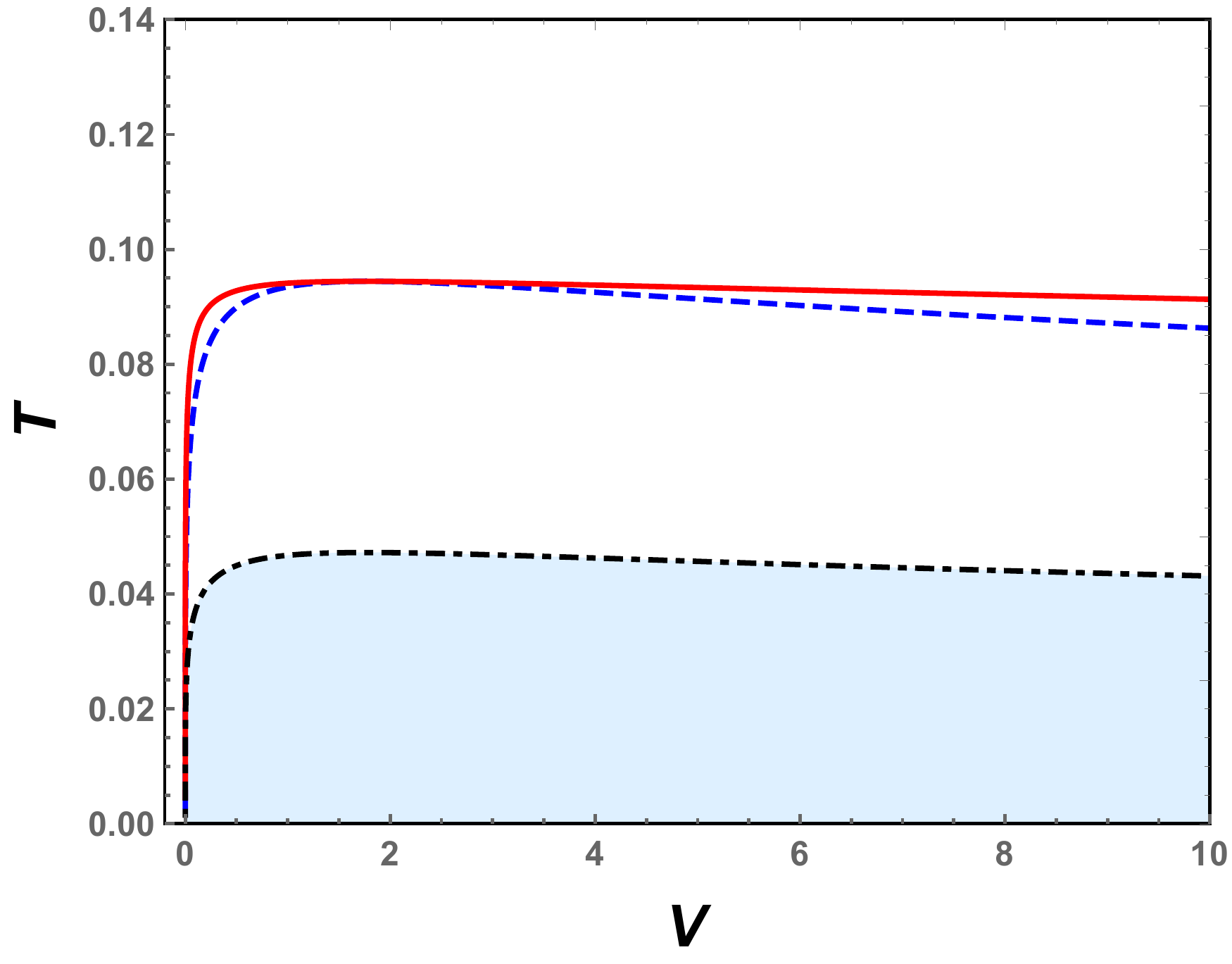}}
	 \subfigure[$\Phi$=0.6]{\label{ctvfour}\includegraphics[width=0.45\textwidth]{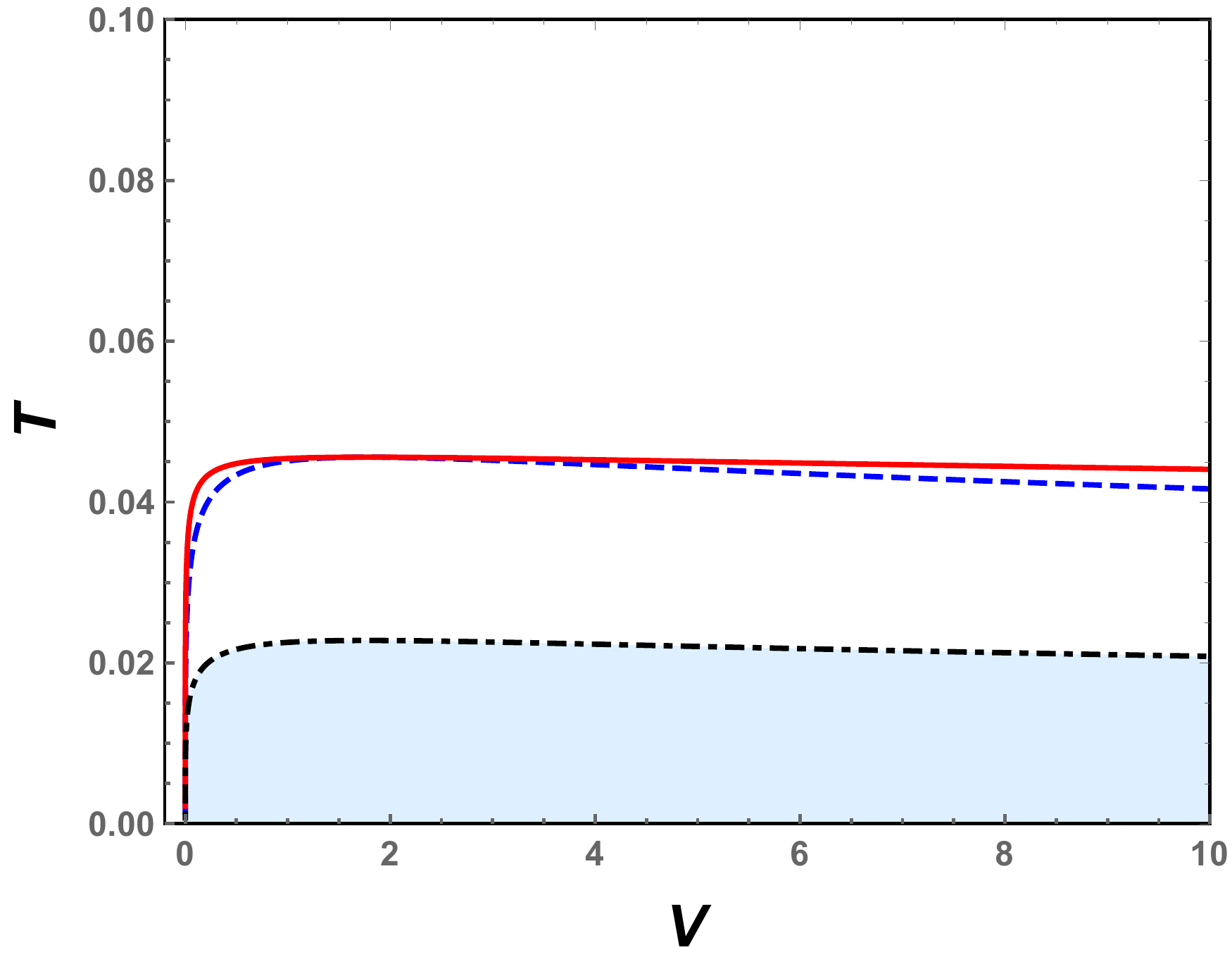}}
	\caption{The coexistence curves (red solid curves), spinodal curves (blue dashed curves), and sign-changing curves (black dot-dashed curves) with $\alpha$=0.1. In the shadow regions, $R_\text{N}$ is positive, otherwise, it is negative. The coexistence region decreases with $\Phi$. When $\Phi$=$\sqrt{3}\pi/{8}$, the coexistence region disappears. \subref{ctvone} $\Phi$=0.1.  \subref{ctvtwo} $\Phi$=0.3. \subref{ctvthree} $\Phi$=0.5. \subref{ctvfour} $\Phi$=0.6.\label{tvcoexistence} }
	
\end{figure}

Next, we will study the normalized scalar curvature in the $T$-$V$ phase diagram and investigate the black hole microstructure for the charged GB-AdS black hole in the grand canonical ensemble. Before examining $R_\text{N}$, we list three characteristic curves
\begin{itemize}
\item Coexistence curve. We have obtained the coexistence curves in the $T$-$V$ plane (see Eq. (\ref{CTV})).
\item Spinodal curve defined by $\partial_V P = 0$. It can be seen from Eq. (\ref{R-CV}) that the normalized scalar curvature diverges at the spinodal curve. Combining with the equation of state (\ref{statep}), the spinodal curve in the $T-V$ plane reads
\begin{eqnarray}
 T_{sp} =\frac{2^{1 / 4} V^{1 / 4}\left(3 \pi^{2}-64 \Phi^{2}\right)}{3 \pi^{5 / 2}\left(\sqrt{2} \sqrt{V}+6 \pi \alpha\right)}.
\end{eqnarray}
\item Sign-changing curve corresponding to $R_\text{N}=0$. This sign-changing curve divides the $T$-$V$ plane into two regions of positive and negative $R_\text{N}$, respectively. Solving $R_\text{N}=0$, we get
\begin{eqnarray}
 T_0 =\frac{T_{sp}}{2}=\frac{V^{1 / 4}\left(3 \pi^{2}-64 \Phi^{2}\right)}{3 \times 2^{3 / 4} \pi^{5 / 2}\left(\sqrt{2} \sqrt{V}+6 \pi \alpha\right)}.
\end{eqnarray}
\end{itemize}
It is clear that the relation $T_{sp}=2 T_0$ holds for different black hole backgrounds \cite{Wei2019,Wei2020a,Kumara2020}. After a simple calculation, we find that if the pressure has a linear relation with the temperature, the relation $T_{sp}=2 T_0$ will hold by using the expression (\ref{R-CV}) of the normalized scalar curvature.

Now we list these three characteristic curves in Fig. \ref{tvcoexistence} for $\Phi$=0.1, 0.3, 0.5, and 0.6. The coexistence, spinodal, and sign-changing curves are, respectively denoted with the red solid, blue dashed, and black dot-dashed curves. The shadow region is for positive $R_\text{N}$ and the other region has negative $R_\text{N}$. Since we do not know whether the equation of state still holds or not in the coexistence regions (below the red curves in Fig. \ref{tvcoexistence}), we exclude them. After this consideration, these regions of positive scalar curvature will be excluded. So for this black hole, only the attractive interaction dominates among the black hole microstructures. This result is similar to that of the five-dimensional neutral GB-AdS black hole \cite{Wei2020a}, and thus the electric potential or the charge does not affect the type of interactions. Moreover, from Fig. \ref{tvcoexistence}, we can find that the critical temperature decreases with $\Phi$, thus the coexistence region shrinks. On the other hand, when $\Phi>\sqrt{3}\pi/{8}$, the small-large black hole phase transition disappears, and the properties of the black hole system get significant change. In the parameter range, the normalized scalar curvature becomes positive, so repulsive interaction dominates among the black hole microstructures.

\begin{figure}
	\center{\includegraphics[width=7cm,height=5cm]{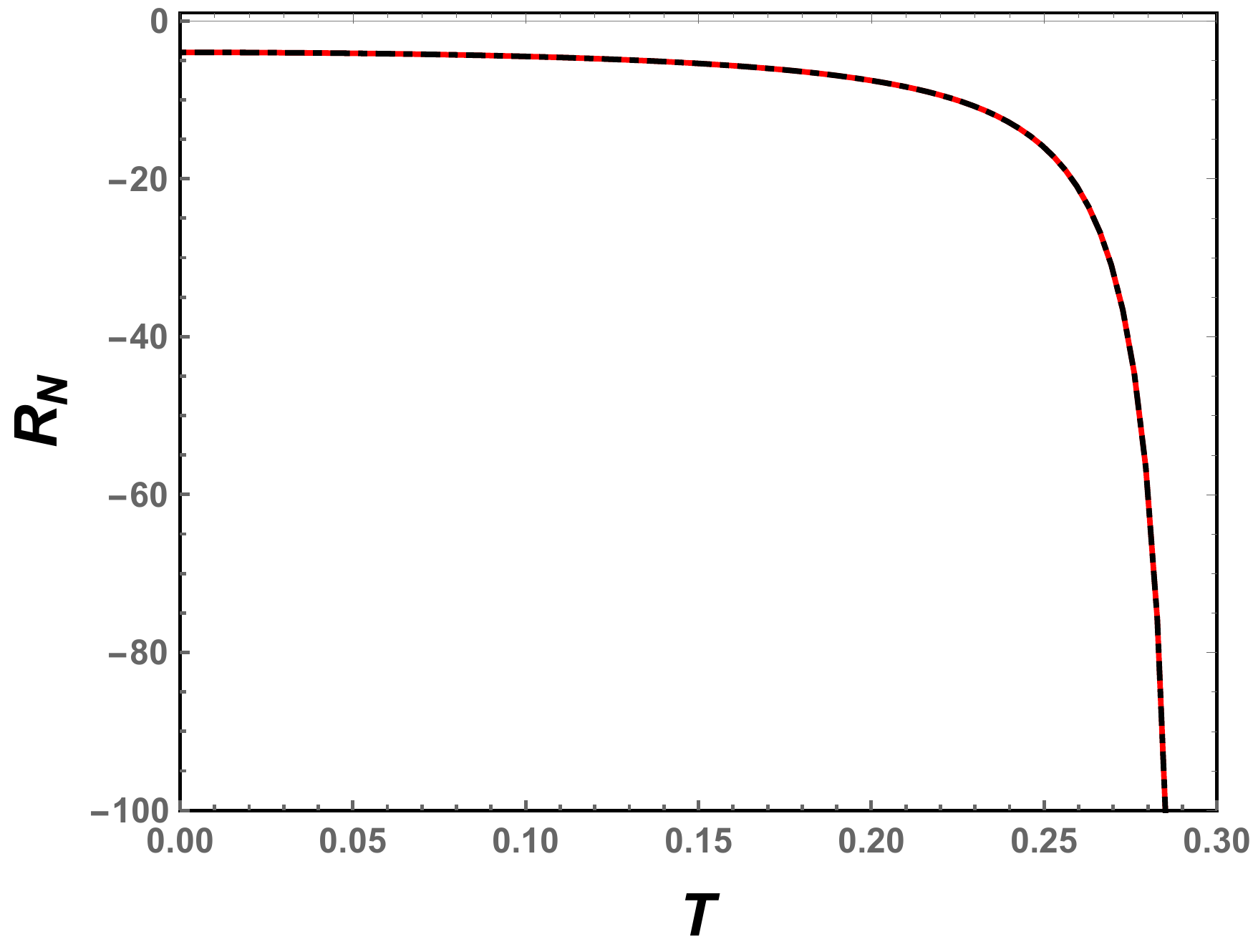}
\caption{Behavior of $R_{N}$ along the coexistence small or large black hole curves for $\Phi$=0.5 and $\alpha$=0.01. Noteh that these two curves coincide with each other.\label{RTVV}}}
\end{figure}

It is also important to examine the behavior of $R_\text{N}$ along the coexistence curve and near the critical point. Substituting Eqs. (\ref{vs}) and (\ref{vl}) into  (\ref{RN}), we find that the normalized scalar curvatures along the coexistence small and large black hole curves are the same
\begin{eqnarray}
R_\text{N}=-\frac{V^{1 / 4}\left(3 \pi^{2}-64 \Phi^{2}\right)\left(-3 \times 2^{1 / 4} \pi^{2} V^{1 / 4}+6 \sqrt{2} \pi^{5 / 2} T \sqrt{V}+36 \pi^{7 / 2} T \alpha+64 \times 2^{1 / 4} V^{1 / 4} \Phi^{2}\right)}{2^{3 / 4}\left(-3 \times 2^{1 / 4} \pi^{2} V^{1 / 4}+3 \sqrt{2} \pi^{5 / 2} T \sqrt{V}+18 \pi^{7 / 2} T \alpha+64 \times 2^{1 / 4} V^{1 / 4} \Phi^{2}\right)^{2}}.\label{CTR}
\end{eqnarray}
The corresponding normalized scalar curvature is plotted in Fig. \ref{RTVV}. When increasing the temperature from zero to its critical values, $R_\text{N}$ starts at a negative value and then decreases with $T$. At the critical temperature, it goes to negative infinity. This behavior of $R_\text{N}$ is consistent with that of the five-dimensional neutral GB-AdS black hole \cite{Wei2020a}. So it seems that the interactions keep unchanged even when the microstructures get a huge change among the black hole phase transition for the charged GB-AdS black hole. This result is also expected to be examined for GB gravity in other dimensions.

\begin{figure}
	\center{\includegraphics[width=7cm,height=5cm]{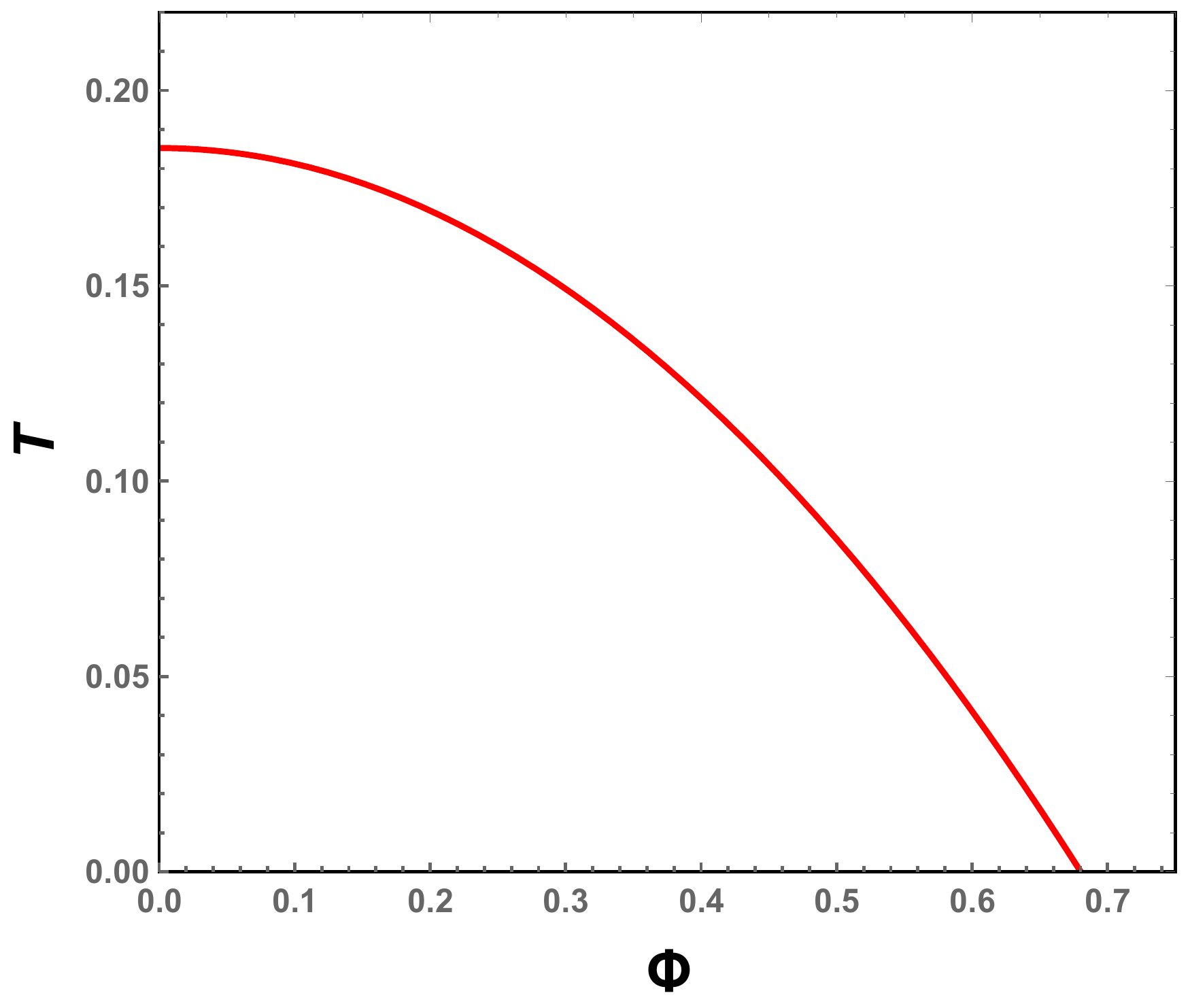}
		\caption{Coexistence curve in the $T$-$\Phi$ plane. We have set $V$ = 0.1 and $\alpha$ = 0.1.\label{T-Phi}}}
\end{figure}

Near the critical point, we expand $R_{N}$ as
\begin{eqnarray}
R_\text{N}=-\frac{\left(3 \pi^{2}-64 \Phi^{2}\right)^{2}}{1728\pi^{6} \alpha }(T_{c}-T)^{-2}+\mathcal{O}(T_{c}-T)^{-1}.
\end{eqnarray}
In the reduced parameter space, it reads
\begin{eqnarray}
R_\text{N}=-\frac{1}{8}(1-\tilde{T})^{-2}-\frac{13}{8}(1-\tilde{T})^{-1}-\frac{27}{32}+\mathcal{O}(1-\tilde{T}),\label{Rtt}
\end{eqnarray}
where $\tilde{T}=\frac{T}{T_{c}}$. That means $R_\text{N}$ has a universal exponent 2 near the critical point. Ignoring the high orders, we obtain the following relation
\begin{eqnarray}
R_\text{N}(1-\tilde{T})^{2}=-\frac{1}{8}.
\end{eqnarray}
This constant is the same as the VdW fluid, four-dimensional charged AdS black hole and five-dimensional neutral GB-AdS black hole \cite{Wei2020a}. Furthermore, the normalized scalar curvature along the coexistence curve can be expressed as a function of the volume $V$. Substituting Eq. (\ref{CTV}) into Eq. (\ref{RN}), we have
\begin{eqnarray}
R_\text{N}=-\frac{4\left(V^{2}-252 \pi^{2} V \alpha^{2}+324 \pi^{4} \alpha^{4}\right)\left(V^{2}-9 \sqrt{2} \pi V^{3 / 2} \alpha-36 \pi^{2} V \alpha^{2}-162 \sqrt{2} \pi^{3} \sqrt{V} \alpha^{3}+324 \pi^{4} \alpha^{4}\right)}{\left(V^{2}-18 \sqrt{2} \pi V^{3 / 2} \alpha+180 \pi^{2} V \alpha^{2}-324 \sqrt{2} \pi^{3} \sqrt{V} \alpha^{3}+324 \pi^{4} \alpha^{4}\right)^{2}},\label{RCC}
\end{eqnarray}
which is independent of the electric potential $\Phi$. For fixed $V$ and $\alpha$, the value of the normalized scalar curvature is uniquely determined, and thus there exists a degeneracy of $\Phi$. On the other hand, we show the coexistence curve in the $T$-$\Phi$ plane as shown in Fig. \ref{T-Phi}. One can see that $T$ decreases with $\Phi$. Meanwhile, the normalized scalar curvature remains unchanged along the coexistence curve. One possible reason is that with the increase of the temperature, the interaction and thermal motion have the same influence on the black hole microstructure.

\section{Conclusions}
\label{Conclusion}

In the present paper, we have analytically studied the phase transition for five-dimensional charged GB-AdS black holes in the grand canonical ensemble.

At first, we constructed the equal area law on each isothermal curve. The analytical coexistence curve was obtained in the $P$-$T$ plane. Based on it, the phase diagrams in the $P$-$V$ and $T$-$V$ planes were investigated. Further, the change of the thermodynamic volume $\Delta V$ among the small-large black hole phase transition was calculated. The results show that $\Delta V$ has a universal exponent of $\frac{1}{2}$ near the critical temperature and pressure. Another interesting result is that in the reduced parameter space, $\Delta V$ has the same expansion behavior as that of the uncharged GB-AdS black hole.

Then we constructed the Ruppeiner geometry for the charged GB-AdS black hole in the grand canonical ensemble. The corresponding normalized scalar curvature was calculated. For small $\Phi$, $R_\text{N}$ is negative, which implies that the attractive interaction dominates among the black hole microstructures. While when $\Phi$ is larger than $\sqrt{3}\pi/8$, the interaction will become repulsive. In the $T$-$V$ phase diagram, we examined the scalar curvature. Three characteristic curves, the coexistence, spinodal, and sign-changing curves were obtained. Employing them, we discussed the type of the interaction in the parameter space. Since the region of positive $R_\text{N}$ always falls in the coexistence region, only attractive interaction dominates among the black hole microstructures. This result is the same as that of the neutral GB-AdS black holes, while different from the charged AdS black holes.

The critical behavior of $R_\text{N}$ was also studied. Near the critical point, $R_\text{N}$ goes to negative infinity, and has a critical exponent 2. Moreover, we observed that $R_\text{N}(1-\tilde{T})^{2}$ equals $-\frac{1}{8}$, which is the same as neutral GB-AdS black holes. So it seems that the charge has no influence on the critical behavior of $R_\text{N}$ for the five-dimensional charged GB-AdS black hole. These results uncover the properties of the black hole microstructure in the grand canonical ensemble. The study is also worth to generalize to other higher dimensional neutral and charged GB-AdS black holes.

\section*{Acknowledgements}
This work was supported by the National Natural Science Foundation of China (Grants No. 11675064 and No. 11875151).


\begin{thebibliography}{99}
	
\bibitem{Hawking1983b}	
S.W. Hawking and D.N. Page,
{\em Thermodynamics of black holes in anti-de Sitter space},
Commun. Math. Phys. \textbf{87}, 577 (1983).

\bibitem{Maldacena}
 J. M. Maldacena,
  {\em The Large N limit of superconformal field theories and supergravity},
    Adv. Theor. Math. Phys. \textbf{2}, 231 (1998), [arXiv:hep-th/9711200].

\bibitem{Gubser}
 S. S. Gubser, I. R. Klebanov, and A. M. Polyakov,
  {\em Gauge theory correlators from noncritical string theory},
     Phys. Lett. B \textbf{428}, 105 (1998), [arXiv:hep-th/9802109].

\bibitem{Witten}
 E. Witten,
  {\em Anti-de Sitter space and holography},
   Adv. Theor. Math. Phys. \textbf{2}, 253 (1998), [arXiv:hep-th/9802150].

\bibitem{Witten2}
 E. Witten,
  {\em Anti-de Sitter space, thermal phase transition, and confinement in gauge theories},
  Adv. Theor. Math. Phys. \textbf{2}, 505 (1998), [arXiv:hep-th/9803131].

\bibitem{Chamblin}
 A. Chamblin, R. Emparan, C. V. Johnson, and R. C. Myers,
  {\em Charged AdS black holes and catastrophic holography},
   Phys. Rev. D \textbf{60}, 064018 (1999), [arXiv:hep-th/9902170].

\bibitem{Chamblin2}
 A. Chamblin, R. Emparan, C. V. Johnson, and R. C. Myers,
  {\em Holography, thermodynamics and fluctuations of charged AdS black holes},
   Phys. Rev. D \textbf{60}, 104026 (1999), [arXiv:hep-th/9904197].

\bibitem{Caldarelli}
 M. M. Caldarelli, G. Cognola, and D. Klemm,
  {\em Thermodynamics of Kerr-Newman-AdS black holes and conformal field theories},
   Class. Quant. Grav. \textbf{17}, 399 (2000), [arXiv:hep-th/9908022].

\bibitem{Henneaux1984}	
M. Henneaux and C. Teitelboim,
{\em The cosmological constant as a canonical variable},
Phys. Lett. B \textbf{143}, 415 (1984).


\bibitem{Teitelboim1985}	
C. Teitelboim,
{\em The cosmological constant as a thermodynamic black hole parameter},
Phys. Lett. B \textbf{158}, 293 (1985).

\bibitem{Creighton1995}
J. D. Creighton and R. B. Mann,
{\em Quasilocal thermodynamics of dilaton gravity coupled to gauge fields},
Phys. Rev. D \textbf{52}, 4569 (1995),
[arXiv:gr-qc/9505007].	

\bibitem{Padmanabhan2002}
T. Padmanabhan,
{\em Classical and quantum thermodynamics of horizons in spherically symmetric space-times},
Class. Quant. Grav. \textbf{19}, 5387 (2002),
[arXiv:gr-qc/0204019].	


\bibitem{Kastor2009d}
D. Kastor, S. Ray, and J. Traschen,
{\em Enthalpy and the Mechanics of AdS Black Holes},
Class. Quant. Grav. \textbf{26}, 195011 (2009),
[arXiv:0904.2765 [hep-th]].	

\bibitem{Dolan00}
B. P. Dolan,
{\em The cosmological constant and black-hole thermodynamic potentials},
Class. Quant. Grav. \textbf{28}, 125020 (2011), [arXiv:1008.5023 [gr-qc]].

\bibitem{Cvetic}
M. Cvetic, G. W. Gibbons, D. Kubiznak, and C. N. Pope,
{\em Black hole enthalpy and an entropy inequality for the thermodynamic volume},
Phys. Rev. D \textbf{84}, 024037 (2011), [arXiv:1012.2888 [hep-th]].

\bibitem{Kubiznak2012b}	
D. Kubiznak and R. B. Mann,
{\em $P$-$V$ criticality of charged AdS black holes},
J. High Energy Phys. \textbf{07}, 033 (2012),
[arXiv:1205.0559 [hep-th]].

\bibitem{Gunasekaran}
S. Gunasekaran, D. Kubiznak, and R. B. Mann,
{\em Extended phase space thermodynamics for charged and rotating black holes and Born-Infeld vacuum polarization},
J. High Energy Phys. \textbf{1211}, 110 (2012), [arXiv:1208.6251 [hep-th]].

\bibitem{Altamirano}
N. Altamirano, D. Kubiznak, and R. B. Mann,
{\em Reentrant Phase Transitions in Rotating AdS Black Holes},
Phys. Rev. D \textbf{88}, 101502 (2013), [arXiv:1306.5756 [hep-th]].

\bibitem{Mann}
N. Altamirano, D. Kubiznak, R. B. Mann, and Z. Sherkatghanad,
{\em Kerr-AdS analogue of triple point and solid/liquid/gas phase transition},
Class. Quant. Grav. \textbf{31}, 042001 (2014), [arXiv:1308.2672 [hep-th]].

\bibitem{Frassino}
A. M. Frassino, D. Kubiznak, R. B. Mann, and F. Simovic,
{\em Multiple reentrant phase transitions and triple points in lovelock thermodynamics},
J. High Energy Phys. \textbf{1409}, 080 (2014), [arXiv:1406.7015 [hep-th]].

\bibitem{Wei0}
S.-W. Wei and Y.-X. Liu,
{\em Triple points and phase diagrams in the extended phase space of charged Gauss-Bonnet black holes in AdS space},
Phys. Rev. D \textbf{90}, 044057 (2014), [arXiv:1402.2837 [hep-th]].

\bibitem{Kostouki}
B. P. Dolan, A. Kostouki, D. Kubiznak, and R. B. Mann,
{\em Isolated critical point from Lovelock gravity},
Class. Quant. Grav. \textbf{31}, 242001 (2014), [arXiv:1407.4783 [hep-th].

\bibitem{Wei1}
S.-W. Wei, P. Cheng, and Y.-X. Liu,
{\em Analytical and exact critical phenomena of $d$-dimensional singly spinning Kerr-AdS black holes},
Phys. Rev. D \textbf{93}, 084015 (2016), [arXiv:1510.00085 [gr-qc]].

\bibitem{Hennigar}
R. A. Hennigar, R. B. Mann, and E. Tjoa,
{\em Superfluid Black Holes},
Phys. Rev. Lett. \textbf{118}, 021301 (2017), [arXiv:1609.02564 [hep-th]].

\bibitem{ZouYue}
M. Zhang, D.-C. Zou, and R.-H. Yue,
{\em Reentrant phase transitions of topological AdS black holes in four-dimensional Born-Infeld-massive gravity},
Adv. High Energy Phys. \textbf{2017}, 3819246 (2017),
[arXiv:1707.04101 [hep-th]].

\bibitem{Hendi3}
S. H. Hendi, G.-Q. Li, J.-X. Mo, S. Panahiyan, and B. E. Panah,
{\em New perspective for black hole thermodynamics in Gauss-Bonnet-Born-Infeld massive gravity},
Eur. Phys. J. C \textbf{76}, 571 (2016),
[arXiv:1608.03148 [gr-qc]].

\bibitem{Hendi4}
S. H. Hendi, R. B. Mann, S. Panahiyan, and B. E. Panah,
{\em van der Waals like behaviour of topological AdS black holes in massive gravity},
Phys. Rev. D \textbf{95}, 021501 (2017),
[arXiv:1702.00432 [gr-qc]].

\bibitem{Momeni}
D. Momeni, M. Faizal, K. Myrzakulov, and R. Myrzakulov,
{\em Fidelity susceptibility as holographic PV-criticality},
Phys. Lett. B \textbf{765}, 154 (2017),
[arXiv:1604.06909 [hep-th]].

\bibitem{Chakraborty}
S. Chakraborty and T. Padmanabhan,
{\em Thermodynamical interpretation of the geometrical variables associated with null surfaces},
Phys. Rev. D \textbf{92}, 104011 (2015),
[arXiv:1508.04060 [gr-qc]].

\bibitem{Weisw}
S.-W. Wei and Y.-X. Liu,
{\em Insight into the Microscopic Structure of an AdS Black Hole from thermodynamic Phase Transition},
Phys. Rev. Lett. \textbf{115}, 111302 (2015), [arXiv:1502.00386 [gr-qc]].


\bibitem{Teo}
D. Kubiznak, R. B. Mann, and M. Teo,
{\em Black hole chemistry: thermodynamics with Lambda},
Class. Quant. Grav. \textbf{34}, 063001 (2017),
[arXiv:1608.06147 [hep-th]].


\bibitem{Vafa}
A. Strominger and C. Vafa,
{\em Microscopic origin of the Bekenstein-Hawking entropy},
Phys. Lett. B \textbf{379}, 99 (1996),
[arXiv:hep-th/9601029]


\bibitem{Lunin}
O. Lunin and S. D. Mathur,
{\em AdS/CFT duality and the black hole information paradox},
Nucl. Phys. B \textbf{623}, 342 (2002),
[arXiv:hep-th/0109154].

\bibitem{Mathur}
O. Lunin and S. D. Mathur,
{\em Statistical interpretation of Bekenstein entropy for systems with a stretched horizon},
Phys. Rev. Lett. \textbf{88}, 211303 (2002),
[arXiv:hep-th/0202072].
	
\bibitem{Cardy1986}	
J. L. Cardy,
{\em OPERATOR CONTENT OF TWO-DIMENSIONAL CONFORMALLY INVARIANT THEORIES},
Nucl. Physics, Sect. B. \textbf{275}, 200 (1986).
	
\bibitem{Banados1992}	
M. Banados, C. Teitelboim, and J. Zanelli,
{\em The Black hole in three-dimensional space-time},
Phys. Rev. Lett. \textbf{69}, 1849 (1992),
[arXiv:hep-th/9204099].

\bibitem{Ruppeiner1979}
G. Ruppeiner,
{\em Thermodynamics: A Riemannian geometric model},
Phys. Rev. A. \textbf{20}, 1608 (1979).

\bibitem{Ruppeiner1995}	
G. Ruppeiner,
{\em Riemannian geometry in thermodynamic fluctuation theory},
Rev. Mod. Phys. \textbf{67}, 605 (1995).

\bibitem{Weiw}
S.-W. Wei and Y.-X. Liu,
{\em Insight into the microscopic structure of an AdS black hole from thermodynamical phase transition},
  Phys. Rev. Lett. \textbf{115}, 111302 (2015), [arXiv:1502.00386
[gr-qc]]; Erratum: Phys. Rev. Lett. \textbf{116}, 169903 (2016).


\bibitem{Wei:2020cqn}
S.-W. Wei and Y.-X. Liu,
{\em New insights into thermodynamics and microstructure of AdS black holes},
Sci. Bull. \textbf{65}, 259 (2020),
[arXiv:2003.00458 [gr-qc]].

\bibitem{Dehyadegari}
 A. Dehyadegari, A. Sheykhi, and A. Montakhab,
   {\em Critical behaviour and microscopic structure of charged AdS black holes via an alternative phase space},
     Phys. Lett. B \textbf{768}, 235 (2017), [arXiv:1607.05333 [gr-qc]].

\bibitem{Zangeneh:2016fhy}
M. K. Zangeneh, A. Dehyadegari, M. R. Mehdizadeh, B. Wang, and A. Sheykhi,
{\em Thermodynamics, phase transitions and Ruppeiner geometry for Einstein-dilaton Lifshitz black holes in the presence of Maxwell and Born-Infeld electrodynamics},
Eur. Phys. J. C \textbf{77}, 423 (2017),
[arXiv:1610.06352 [hep-th]].

\bibitem{Moumni}
 M. Chabab, H. E. Moumni, S. Iraoui, K. Masmar, and S. Zhizeh,
  {\em More insight into microscopic properties of RN-AdS black hole surrounded by quintessence via an alternative extended phase space},
     Int. J. Geom. Meth. Mod. Phys. \textbf{15}, 1850171 (2018), [arXiv:1704.07720 [gr-qc]].

\bibitem{Deng}
 G.-M. Deng and Y.-C. Huang,
   {\em $Q$-$\Phi$ criticality and microstructure of charged AdS black holes in f($R$) gravity},
     Int. J. Mod. Phys. A \textbf{32}, 1750204 (2017), [arXiv:1705.04923 [gr-qc]].

\bibitem{Sheykhi}
 M. K. Zangeneh, A. Dehyadegari, A. Sheykhi, and R. B. Mann,
   {\em Microscopic origin of black hole reentrant phase transitions},
     Phys. Rev. D \textbf{97}, 084054 (2018), [arXiv:1709.04432 [hep-th]].

\bibitem{Miao}
 Y.-G. Miao and Z.-M. Xu,
   {\em Microscopic structures and thermal stability of black holes conformally coupled to scalar fields in five dimensions},
  Nucl. Phys. B \textbf{942}, 205 (2019),
   [arXiv:1711.01757 [hep-th]].

\bibitem{Miao2}
 Y.-G. Miao and Z.-M. Xu,
   {\em Thermal Molecular Potential among Micromolecules in Charged AdS Black Holes},
  Phys. Rev. D \textbf{98}, 044001 (2018),
   [arXiv:1712.00545 [hep-th]].

\bibitem{Miao3}
Y.-G. Miao and Z.-M. Xu,
{\em Interaction Potential and Thermo-correction to the Equation of State for Thermally Stable Schwarzschild AdS Black Holes},
Sci. China-Phys. Mech. Astron. \textbf{62}, 010412 (2019),
[arXiv: 1804.01743 [hep-th]].

\bibitem{Miao2018}
Y.-G. Miao and Z.-M. Xu,
{\em Parametric phase transition for a Gauss-Bonnet AdS black hole},
Phys. Rev. D \textbf{98}, 084051 (2018),
[arXiv:1806.10393 [hep-th]].

\bibitem{Li}
 D.-D. Li, S.-S. Li, L.-Q. Mi, and Z.-H. Li,
   {\em Insight into black hole phase transition from parametric solutions},
  Phys. Rev. D \textbf{96}, 124015 (2017).

\bibitem{Yang:2018ixs}
S.-J. Yang, Y.-B. He, and J.-J. Du,
{\em The first law and Ruppeiner geometry for Grumiller black hole},
EPL {\bf 121}, 50005 (2018).

\bibitem{Chen}
 Y. Chen, H. Li, and S.-J. Zhang,
   {\em Microscopic explanation for black hole phase transitions via Ruppeiner geometry: two competing mechanisms},
   [arXiv:1812.11765 [hep-th]].

\bibitem{Guo}
 X.-Y. Guo, H.-F. Li, L.-C. Zhang, and R. Zhao,
   {\em Microstructure and continuous phase transition of RN-AdS black hole},
   Phys. Rev. D \textbf{100}, 064036 (2019),
   [arXiv:1901.04703 [gr-qc]].

\bibitem{Du}
 Y.-Z. Du, R. Zhao, and L.-C. Zhang,
  {\em Microstructure and continuous phase transition of the gauss-bonnet ads black hole},
   [arXiv:1901.07932 [hep-th]].

\bibitem{Sheykhi:2019vzb}
A. Sheykhi, M. Arab, Z. Dayyani, and A. Dehyadegari,
{\em Alternative approach towards critical behavior and microscopic structure of the higher dimensional Power-Maxwell black holes},
Phys. Rev. D \textbf{101}, 064019 (2020),
[arXiv:1909.11445 [physics.gen-ph]].

\bibitem{Xuz}
 Z.-M. Xu, B. Wu, and W.-L. Yang,
  {\em The fine micro-thermal structures for the Reissner-Nordstrom black hole},
   [arXiv:1910.03378 [gr-qc]].		

\bibitem{GhoshBhamidipati}
A. Ghosh and C. Bhamidipati,
{\em Thermodynamic geometry for charged Gauss-Bonnet black holes in AdS spacetimes},
Phys. Rev. D \textbf{101}, 046005 (2020)
[arXiv:1911.06280 [gr-qc]].
	
\bibitem{Wei2019}
S.-W. Wei, Y.-X. Liu, and R. B. Mann,
{\em Repulsive Interactions and Universal Properties of Charged Anti-de Sitter Black Hole Microstructures},
Phys. Rev. Lett. \textbf{123}, 071103 (2019),
[arXiv:1906.10840 [gr-qc]].

\bibitem{Wei2019d}
S.-W. Wei, Y.-X. Liu, and R. B. Mann,
{\em Ruppeiner Geometry, Phase Transitions, and the Microstructure of Charged AdS Black Holes},
Phys. Rev. D \textbf{100}, 124033 (2019),
[arXiv:1909.03887 [gr-qc]].

\bibitem{Kumara2020}
A. N. Kumara, C. A. Rizwan, K. Hegde, and Ajith K. M,
{\em Repulsive Interactions in the Microstructure of Regular Hayward Black Hole in Anti-de Sitter Spacetime},
[arXiv:2003.10175 [gr-qc]].

\bibitem{Bairagya:2020dtl}
J. D. Bairagya, K. Pal, K. Pal, and T. Sarkar,
{\em Geometry of AdS black hole thermodynamics in extended phase space},
[arXiv:2004.06498 [hep-th]].

\bibitem{Yerra:2020oph}
P. K. Yerra and C. Bhamidipati,
{\em Ruppeiner Geometry, Phase Transitions and Microstructures of Black Holes in Massive Gravity},
[arXiv:2006.07775 [hep-th]].


\bibitem{Wu:2020fij}
B. Wu, C. Wang, Z.-M. Xu, and W.-L. Yang,
{\em Ruppeiner geometry and thermodynamic phase transition of the black hole in massive gravity},
[arXiv:2006.09021 [gr-qc]].

\bibitem{Vaid}
C. L. A. Rizwan, A. N. Kumara, K. Hegde, and D. Vaid,
{\em Coexistent Physics and Microstructure of the Regular Bardeen Black Hole in Anti-de Sitter Spacetime},
[arXiv:2008.06472 [gr-qc]].

\bibitem{Rizwan}
A. N. Kumara, C.L. A. Rizwan, K. Hegde, M. S. Ali, and Ajith K.M,
{\em Ruppeiner Geometry, Reentrant Phase transition and Microstructure of Born-Infeld AdS Black Hole},
[arXiv:2007.07861 [gr-qc]].

\bibitem{Mansoori}
S. A. H. Mansoori, M. Rafiee, and S.-W. Wei,
{\em Universal criticality of thermodynamic curvatures for charged AdS black holes},
[arXiv:2007.03255 [gr-qc]].

\bibitem{Kumara}
A. N. Kumara, C.L. A. Rizwan, K. Hegde, M. S. Ali, and Ajith K.M,
{\em Microstructure of five-dimensional neutral Gauss-Bonnet black hole in anti-de Sitter spacetime via $P$-$V$ criticality},
[arXiv:2006.13907 [gr-qc]].

\bibitem{Mannw}
S.-W. Wei, Y.-X. Liu, and R. B. Mann,
{\em Novel dual relation and constant in Hawking-Page phase transition},
[arXiv:2006.11503 [gr-qc]].

\bibitem{Dehyadegariw}
A. Dehyadegari, A. Sheykhi, and S.-W. Wei,
{\em Microstructure of charged AdS black hole via $P$-$V$ criticality},
[arXiv:2006.12265 [gr-qc]].

\bibitem{Dehyadegari:2020ebz}
A. Dehyadegari, A. Sheykhi, and S.-W. Wei,
{\em Microstructure of charged AdS black hole via $P$-$V$ criticality},
[arXiv:2006.12265 [gr-qc]].

\bibitem{Wei2020d}
S.-W. Wei and Y.-X. Liu,
{\em Extended thermodynamics and microstructures of four-dimensional charged Gauss-Bonnet black hole in AdS space},
 Phys. Rev. D \textbf{101}, 104018 (2020),
[arXiv:2003.14275 [gr-qc]].

\bibitem{Wei2020a}
S.-W. Wei and Y.-X. Liu,
{\em Intriguing microstructures of five-dimensional neutral Gauss-Bonnet AdS black hole},
Phys. Lett. B \textbf{803}, 135287 (2020),
[arXiv:1910.04528 [gr-qc]].	
	
\bibitem{Cai2013}	
R.-G. Cai, L.-M. Cao, L. Li, and R.-Q. Yang,
{\em $P$-$V$ criticality in the extended phase space of Gauss-Bonnet black holes in AdS space},
J. High Energy Phys. \textbf{09}, 005 (2013),
[arXiv:1306.6233 [gr-qc]].
	
\bibitem{Zou2014c}
D.-C. Zou, Y. Liu, and B. Wang,
{\em Critical behavior of charged Gauss-Bonnet AdS black holes in the grand canonical ensemble},
Phys. Rev. D \textbf{90}, 044063 (2014),
[arXiv:1404.5194 [hep-th]].

\bibitem{Belhaj2015}
A. Belhaj, M. Chabab, H. E. moumni, K. Masmar, and M. Sedra,
{\em Maxwell's equal-area law for Gauss-Bonnet-Anti-de Sitter black holes},
Eur. Phys. J. C \textbf{75}, 71 (2015),
[arXiv:1412.2162 [hep-th]].

\bibitem{Wei2015c}
S.-W. Wei and Y.-X. Liu,
{\em Clapeyron equations and fitting formula of the coexistence curve in the extended phase space of charged AdS black holes},
Phys. Rev. D \textbf{91}, 044018 (2015),
[arXiv:1411.5749 [hep-th]].

\bibitem{Ruppeiner2014b}
G. Ruppeiner,
{\em Thermodynamic curvature and black holes},
Springer Proc. Phys. \textbf{153}, 179 (2014),
[arXiv:1309.0901 [gr-qc]].

\bibitem{Ruppeiner2010}
G. Ruppeiner,
 {\em Thermodynamic curvature measures interactions},
  Am. J. Phys. \textbf{78}, 1170 (2010),
[arXiv:1007.2160 [cond-mat.stat-mech]]

\bibitem{Landau1977}
 L. D. Landau and E. M. Lifshitz,
 {\em Statistical Physics} (Pergamon, New York, 1977).

\end{thebibliography}
\end{document}